\documentclass{nature}
\usepackage[T1]{fontenc}
\usepackage{graphicx}
\usepackage{pdflscape}
\usepackage{threeparttable}
\usepackage{xcolor}
\usepackage{ulem}
\usepackage{float}
\restylefloat{table}

\usepackage{caption}
\usepackage[caption = false]{subfig}
\usepackage{academicons}
\usepackage{amsmath, amssymb}
\usepackage{tabularx}

\definecolor{orcidlogocol}{HTML}{A6CE39}

\makeatletter

\let\saved@includegraphics\includegraphics
\AtBeginDocument{\let\includegraphics\saved@includegraphics}
\makeatother

\definecolor{dkblue}{RGB}{54, 86, 169}

\def\be{\begin{eqnarray}}
\def\ee{\end{eqnarray}}

\makeatletter

\let\saved@includegraphics\includegraphics
\AtBeginDocument{\let\includegraphics\saved@includegraphics}
\renewenvironment*{figure}{\@float{figure}}{\end@float}
\makeatother

\makeatletter
\def\@fnsymbol#1{\ensuremath{\ifcase#1\or \dagger\or \ddagger\or
 \mathsection\or \mathparagraph\or \|\or **\or \dagger\dagger
 \or \ddagger\ddagger \else\@ctrerr\fi}}
\makeatother

\newcommand{\EXTTAB}[1] {Extended Data Table~\ref{#1}}
\newcommand{\EXTFIG}[1] {Extended Data Figure~\ref{#1}}

\usepackage{color}
\usepackage{hyperref}
\usepackage{graphicx}
\usepackage{longtable}
\usepackage{supertabular}
\usepackage{multibib}
\newcites{methods}{Methods References}

\title{An assessment of the
Association Between a Fast Radio Burst and Binary Neutron Star Merger}

\author{Alexandra Moroianu$^{*1, 2}$, Linqing Wen$^{*1, 2}$, Clancy W. James$^{3}$, Shunke Ai$^{4, 5}$, Manoj Kovalam$^{1, 2}$, Fiona Panther$^{1, 2}$, Bing Zhang$^{4, 5}$}

\begin{document}

\maketitle
\begin{affiliations}
 \item Australian Research Council Centre of Excellence for Gravitational Wave Discovery (OzGrav)
 \item Department of Physics, University of Western Australia, Crawley WA 6009, Australia
 \item International Centre for Radio Astronomy Research, Curtin University, Bentley, WA 6102, Australia
 \item Nevada Center for Astrophysics, University of Nevada, Las Vegas, NV 89154, USA
 \item Department of Physics and Astronomy, University of Nevada, Las Vegas, NV 89154, USA
 \end{affiliations}

\bigskip

\begin{abstract} 
\bf Fast radio bursts (FRBs) are mysterious bright millisecond-duration radio bursts at cosmological distances\cite{lorimer07,chime2021}. While young magnetars have been put forward as the leading source candidate\cite{Michilli2018_121102,chime2020,bochenek2020,li21c,mereghetti20},
recent observations suggest there may be multiple FRB progenitor classes\cite{chime2021,Kirsten2021_M81GC}.
It has long been theorised that FRBs could be emitted from compact object mergers\cite{platts2019} --- cataclysmic events such as binary neutron star (BNS) mergers that may be detectable in gravitational waves (GWs) by the ground-based Laser Interferometer Gravitational Wave Observatory (LIGO)\cite{aligo2015} and Virgo\cite{acernese2014}. Here we report a potential coincidence between the only BNS merger event GW190425\cite{abbott2020} out of 21 GW sources detected during the first six months of LIGO-Virgo's 3rd Science Run and a bright, non-repeating FRB event, FRB 20190425A\cite{chime2021}, from a search using public GW and CHIME FRB data. The FRB is located within the GW's sky localization area, occurred 2.5 hours after the GW event, and has a dispersion measure consistent with the distance inferred from GW parameter estimation\cite{abbott2021}. The chance probability of a coincidence between unrelated FRB and GW events in the databases is estimated to be $0.0052$ ($2.8\,\sigma$).
We estimate the chance of CHIME detecting such an event to range from 0.4\% for a beam-centre detection to 68\% if a bright burst is detectable in a far sidelobe. This potential association is consistent with the theory\cite{zhang2014} that the BNS merger leaves behind a supramassive, highly magnetized compact object, which collapses to form a black hole after losing angular momentum due to spindown and makes an FRB through ejecting the magnetosphere\cite{falcke2014}. If such a physical association is established, the equation of state of the post-merger compact object is likely stiff, with a Tolman-Oppenheimer-Volkoff non-spinning maximum mass\cite{tov1939} 
$M_{\rm TOV} > 2.63_{-0.23}^{+0.39} M_\odot$ for a neutron star remnant, or $M_{\rm TOV} > 2.31_{-0.08}^{+0.24} M_\odot$ for a quark star remnant.
\end{abstract}

To date, more than 600 FRBs have been detected at radio frequencies between $110\,\mathrm{MHz}$ and $8\,\mathrm{GHz}$\cite{pleunis2021,gajjar2018}. The high all-sky rate\cite{chime2021} and event rate density\cite{ravi2019,luo2020} of FRBs, combined with the fact that some FRB sources emit repeated bursts\cite{spitler2016,chime2021}, suggest that the majority of FRBs are not produced from cataclysmic channels such as compact object mergers.
However, a small sub-population of FRBs associated with cataclysmic events would be difficult to detect in these analyses\cite{chime2021}, and be consistent with cataclysmic event rates \cite{ravi2019}.
Furthermore, extensive follow-up studies have failed to detect repeating radiation from some nearby FRBs\cite{lorimer07,James2020a}. There exist several theories that predict the association of an FRB with a GW event due to compact binary coalescence between two neutron stars, a neutron star and a black hole, or even two charged black holes\cite{platts2019}. In particular, binary neutron star (BNS) mergers have long been theorised to emit FRB-like signals before\cite{piro2012,zhang2020b}, during\cite{totani2013} or after\cite{zhang2014} the merger. With the publicly available GW catalogue GWTC-2\cite{abbott2021}
and the newly released first FRB catalogue\cite{chime2021} from the Canadian Hydrogen Intensity Mapping Experiment FRB project (CHIME/FRB), it is possible to test these theories by searching for GW-FRB associations. 
We conduct a search for GW-FRB coincidences using CHIME/FRB's first FRB catalogue containing 535 new FRB sources\cite{chime2021} (observed July 2018 - July 2019), 171 of which overlap with the first half of LIGO's $3^{\rm rd}$ Science Run (O3a: April 1, 2019 - October 1, 2019)\cite{abbott2021}. Our search time window is chosen to be asymmetrical and 26 hours wide, encompassing FRBs that occur up to 2 hours before a GW signal\cite{totani2013} and 24 hours after\cite{zhang2014} (see Methods). A GW-FRB pair is considered to be coincident in time if an FRB falls within the time window of a GW signal. CHIME/FRB localization\cite{chime2021} is accurate on the order of arcminutes, more precise than even the best GW localization of tens of square degrees\cite{abbott2020b}. Therefore, we consider a GW-FRB pair spatially coincident if the FRB lies within the 90\text{\%} credible interval\cite{abbott2021} of a candidate GW's localization (see Methods).

\begin{figure}
 \centering
 \includegraphics[width=\textwidth]{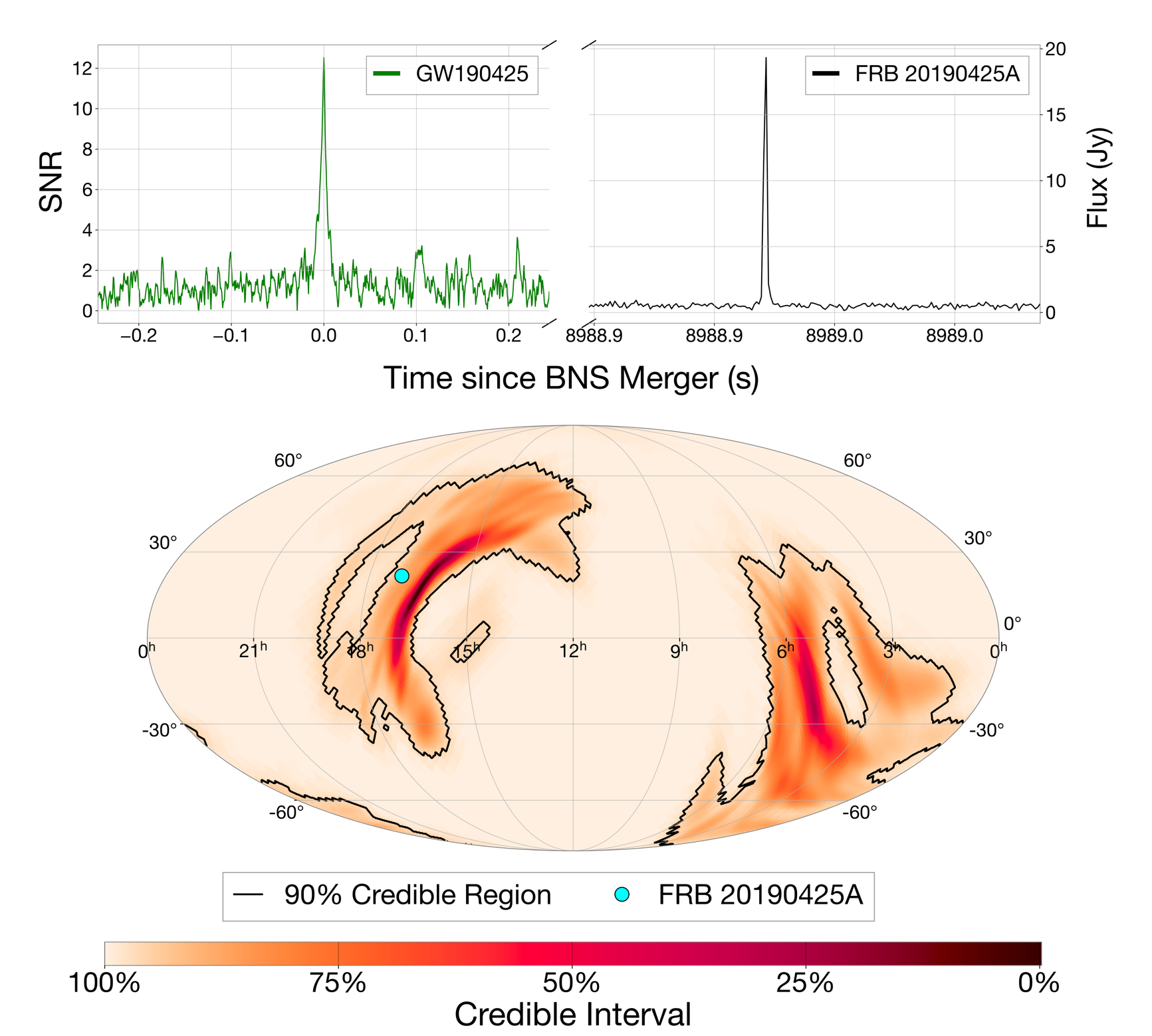}
 \caption{{\bf Temporal and spatial coincidence of GW190425 and FRB 20190425A}. Top left: LIGO Livingston signal to noise ratio (SNR) of GW190425 (see Methods). Top right: flux (in Jy) of FRB 20190425A\cite{chime2021}, occurring $2.5\ \rm hours$ after the merger. Bottom: Sky direction of FRB 20190425A (cyan) and the localization of GW190425\cite{abbott2021} (90\% contour indicated by black line).}

 \label{fig:skymap}
\end{figure}

We find an apparently non-repeating FRB 20190425A\cite{chime2021} temporally and spatially coincident with a GW merger event: GW190425\cite{abbott2020, abbott2021}. GW190425 was observed on April 25, 2019 08:18:05 UTC by LIGO Livingston (LIGO Hanford was offline) with a false alarm rate of $\rm FAR = 1/69,000\ \mathrm{yr^{-1}}$. There was no significant detection made by Virgo due to its lower sensitivity, which helps constrain the sky localization of the event. GW190425 is a BNS merger event, however the remnant mass of the system in the source frame, $M_{\rm tot}^{\rm final} = 3.23^{+0.33}_{-0.11}\ M_{\odot}$, is significantly larger than that predicted by known Galactic BNS systems\cite{abbott2021}. FRB 20190425A is a curiously bright (see Methods) radio transient with fluence $31.6 \pm 4.2\ \mathrm{Jy\ ms}$ (assuming a beam-centre detection) and burst width $3.799 \pm 0.002 \cdot 10^{-4}\ \mathrm{s}$.
It has a broadband emission across CHIME's 400--800\,MHz bandwidth, and a single-peaked morphology, which is observed in 30\% of the CHIME FRB population, and is associated with non-repeating FRBs.
In contrast, repeating FRBs typically exhibit narrow-band structure\cite{chime2021}. It also has an unusually low dispersion measure (DM) of $128.2\ \mathrm{pc\ cm^{-3}}$, indicating an origin within $z < 0.0394$, assuming DM originates only from the Milky Way and the intergalactic medium\cite{macquart2020}.
It was observed 2.5 hours after GW190425 at 10:46:33 UTC at a peak frequency of 591.8 MHz, and localized to J2000 celestial coordinates RA = $255.72 \pm 0.14 ^{\circ}$, DEC = $21.52 \pm 0.18 ^{\circ}$. This places it in the 66.7\% credible interval of GW190425's refined GWTC-2 skymap\cite{abbott2021} (Figure \ref{fig:skymap}). Parameter estimation\cite{abbott2021} localized the BNS merger event to redshift $z = 0.03^{+0.01}_{-0.02}$. This is within the upper limit ($z < 0.0394$) specified by FRB 20190425A's DM (see Methods), making the two signals coincident within the error margins of their distances.

\begin{figure}
    \centering
    \includegraphics[width=0.5\textwidth]{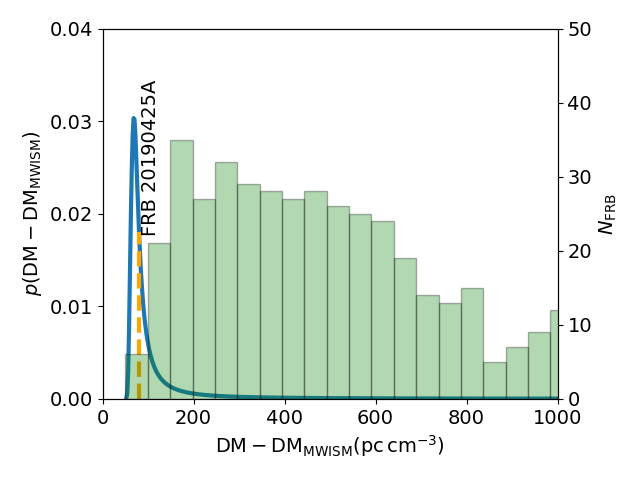}
    \caption{{\bf The expected dispersion measure (DM) distribution inferred from GW190425, compared to non-repeating CHIME FRBs.} Blue solid line: the probability distribution of DM after subtracting the contribution from the Milky Way's interstellar medium (DM-DM$_{\rm MWISM}$; see Methods). Orange dashed line: the observed value for FRB 20190425A, compared to the number of non-repeating CHIME FRBs ($N_{\rm FRB}$, green histogram).
    }
    \label{fig:expected_DM}
\end{figure}

To quantify the likelihood of
this association being entirely coincidental, we estimate the 
probability of chance coincidence of GW190425 and FRB20190425A assuming 
that FRB and GW detections are independent. Evidence against this hypothesis comes from three sources: the spatial, temporal, and DM coincidence. We exclude repeating FRBs in this analysis (see Methods). The likelihood of a chance temporal coincidence between GW190425 and FRB 20190425A ($P_{\rm T}$) is taken to be the probability of CHIME randomly detecting an FRB in a 2.5 hour stretch of time. We estimate a CHIME FRB detection rate of 1.93 per day around the time of GW190425, leading to a Poisson probability of $P_{\rm T}=0.18$. We define $P_{\rm S}$ to be the probability that a random CHIME FRB would be detected during this 2.5 hour interval with equal to or higher likelihood in the GW190425 skymap than FRB 20190425A (i.e.\ within the 66.7\% credible interval). Using the estimated declination dependence of CHIME's exposure\cite{chime2021}, we estimate $P_{\rm S}=0.265$ (see Methods). If we instead use our initial temporal (26\,hr time window) and spatial (90\% credible interval) selection criteria, we find 0.88 ($P_{\rm T}$) and 0.15 ($P_{\rm S}$) respectively. The probability of an FRB originating at redshift $z$ having a dispersion measure DM has been estimated using a sample of localized FRBs\cite{macquart2020}. We produce the expected DM probability distribution for the published redshift posterior distribution\cite{abbott2020} of GW190425. We find that FRB 20190425A lies in the 46\% credible interval of GW190425's DM probability distribution. Only one of 473 other FRBs exhibits a DM with a more-likely value (Figure~\ref{fig:expected_DM}). We thus find $P_{\rm DM}$, the chance that a random CHIME FRB would have a DM with an equal or better match to expectations, to be $2/474 \approx 0.004$.

CHIME has shown that the temporal, spatial and DM distributions of FRBs are independent\cite{CHIME2021_lat_dependence}, except for a potential correlation\cite{CHIME2021_structure} between FRBs with DM $\sim$ 800\,pc\,cm$^{-3}$ and large-scale structure at $z\sim0.4$.
We use $P_{\rm T} P_{\rm S} P_{\rm DM} = 0.18 \cdot 0.265 \cdot 0.004 = 1.9 \cdot 10^{-4}$ as evidence against the null hypothesis of a chance coincidence.
The chance probability of finding an FRB with smaller product $P_{\rm T} P_{\rm S} P_{\rm DM}$, i.e.\ the p-value, is $0.0052$ (see Methods).
Note the close proximity of this FRB in time, sky direction and distance, as implied by DM measurements, to GW190425 makes this association more significant than the random chance of finding any FRBs within the search windows used for initial discovery (13.5\% probability).
Of further interest, GW190425-FRB 20190425A is the only GW-FRB pair that survived our time-spatial coincidence criteria and it is linked to the only BNS event out of the 21 GW sources detected. This encourages us to consider a potential astrophysical association between GW190425 and FRB 20191425A.

\begin{figure}
    \centering
    \includegraphics[width=\textwidth]{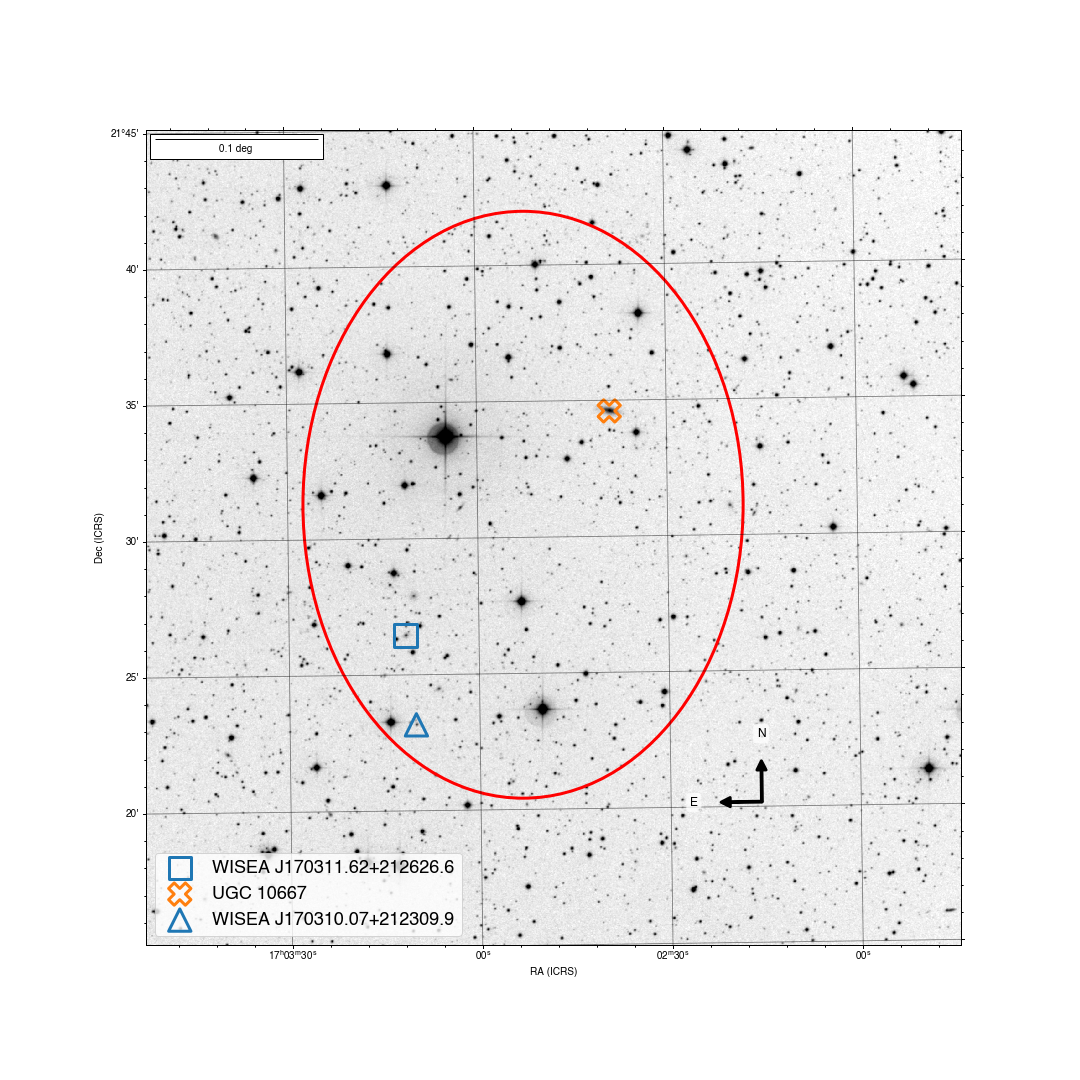}
    \caption{{\bf Potential host association for GW190425-FRB 20190425A.} We investigate hosts within an error ellipse defined by the 1-$\sigma$ uncertainties in RA ($\pm0.14^{\circ}$) and DEC ($\pm 0.18^{\circ}$) centered on the optimal FRB location derived by CHIME\cite{chime2021} (red ellipse). Redshifts of the identified extragalactic objects are listed in Methods. UGC 10667 (orange cross) is the only host with a redshift consistent with the redshift derived for the GW source. Background image: DSS2-red (\url{http://archive.eso.org/dss/dss}).}
    \label{fig:host-assoc}
\end{figure}

We search for a host galaxy inside the reported error ellipse of FRB 20190425A's central localization (see Methods). We identify one candidate in the NASA Extragalactic Database (NED)\footnote{The NASA/IPAC Extragalactic Database (NED) is funded by the National Aeronautics and Space Administration and operated by the California Institute of Technology.} within a redshift range $0.001 < z < 0.08$ consistent with the BNS merger redshift range: UGC 10667 (Figure \ref{fig:host-assoc}). This galaxy has a redshift consistent with the BNS merger and the FRB DM. UGC 10667 has an offset of 5.05 arcminutes from the optimal FRB location and a redshift of $z=0.03136\pm 0.00009$. The expected number of galaxies\cite{Gehrels2016} within the search volume is $N_\mathrm{gals}\sim0.173$ (see Methods) assuming a galaxy density of $\rho_\mathrm{gals}=2.35 \cdot 10^{-3}\,\mathrm{Mpc}^{-3}$ and a maximum
luminosity distance of $d_\mathrm{L} = 255.85\,\mathrm{Mpc}$. Therefore, to investigate the association, we encourage follow-up observations of UGC 10667 to search for evidence of the merger event, such as the afterglow emission of the ejecta in the broad band, especially in the radio band at sub-GHz frequencies\cite{nakar11,gao13}.

Though we cannot definitively assign the potential GW-FRB association to a single theory, it is consistent with the GW, short gamma-ray burst (sGRB) and FRB association theory invoking the collapse of a post-BNS-merger magnetar\cite{zhang2014}. The FRB generation mechanism is the so-called ``blitzar'' mechanism\cite{falcke2014}, which has been confirmed through numerical simulations\cite{most2018}. Within this scenario, the 2.5 hour delay time between the FRB and the GW event is the survival time of the supramassive neutron star before collapsing into a black hole\cite{zhang2014}, which is consistent with the expected delay timescale range for a supramassive magnetar both in theory and in observational data (see Methods). The chance of CHIME detecting such a single burst within its $1.3^{\circ}$--$2.5^{\circ}$ wide primary beam is 0.4--0.8\%, increasing to 14\% when allowing for a detection in the far sidelobes (see Methods).
The duration of the FRB is of the order of light crossing time of the ejected magnetosphere of the collapsing supramassive neutron star, which is of the order of its spin period. This is consistent with the $\sim$ millisecond duration of FRB 20190425A.
As an initially rapidly rotating supramassive neutron star collapses to a black hole after losing angular momentum, the total ejected electromagnetic (EM) energy\cite{zhang2014,most2018} is $E_{\rm EM} \sim 1.7\cdot10^{44} \ {\rm erg} \ (B/10^{14} \,\mathrm{Gauss})^2 (R_{\rm NS} / 10 \,{\rm km})^3$, where $B$ and $R_{\rm NS}$ correspond to the surface magnetic field and the radius of the neutron star, respectively. We calculate the FRB's total isotropic radio emission energy as $E_\mathrm{FRB,iso} = 3.72\cdot10^{38}\,\mathrm{erg}$ using the measured fluence of FRB 20190425A ($3.2 \cdot 10^{-28}\,\mathrm{J\,m^{-2}} \cdot 400\,\mathrm{MHz}$) and the most probable luminosity distance ($156\,\mathrm{Mpc}$) of GW190425. The true energy of the emitter should be $E_\mathrm{FRB,iso} f_b \eta_r^{-1}$, where $f_b\leq1$ is the beaming factor and $\eta_r$ is the radio emission efficiency. For the blitzar FRB model, the EM energy injection is essentially isotropic\cite{falcke2014,most2018}, so $f_b \sim 1$. This implies $\eta_r\sim2\cdot10^{-6}$, similar to the reported value for the Galactic FRB 200428 (ref.\ \cite{chime2020, bochenek2020,li21c,mereghetti20}). BNS mergers eject dense neutron-rich material that could prevent the escape of coherent radio emission. For an FRB to be observed following the merger, the line of sight needs to be cleared by a sGRB jet\cite{zhang2014}. However, the sGRB does not need to be bright enough to be detected by current gamma-ray telescopes. For example, a sGRB with the brightness of GRB 170817A would be able to clear the ejecta and escape detection at the distance of $156\,\mathrm{Mpc}$ (ref.\ \cite{abbott2017b,zhangbb2018}). We note that an excess of gamma-rays was detected 0.5--5.9\,s after GW190425 by the \textit{International
Gamma-Ray Astrophysics Laboratory} (\textit{INTEGRAL}) using the SPI-ACS subsystem. This was reported as a candidate sGRB of marginal significance associated with GW190425 by two independent analyses\cite{INTEGRALGCN2, pozanenko2019}. Both the localization of the sGRB reported by \textit{INTEGRAL}, and the non-detection by \textit{FERMI},\cite{pozanenko2019} is consistent with the FRB and GW association.

\begin{figure}
\centering
\includegraphics[width=0.8\textwidth]{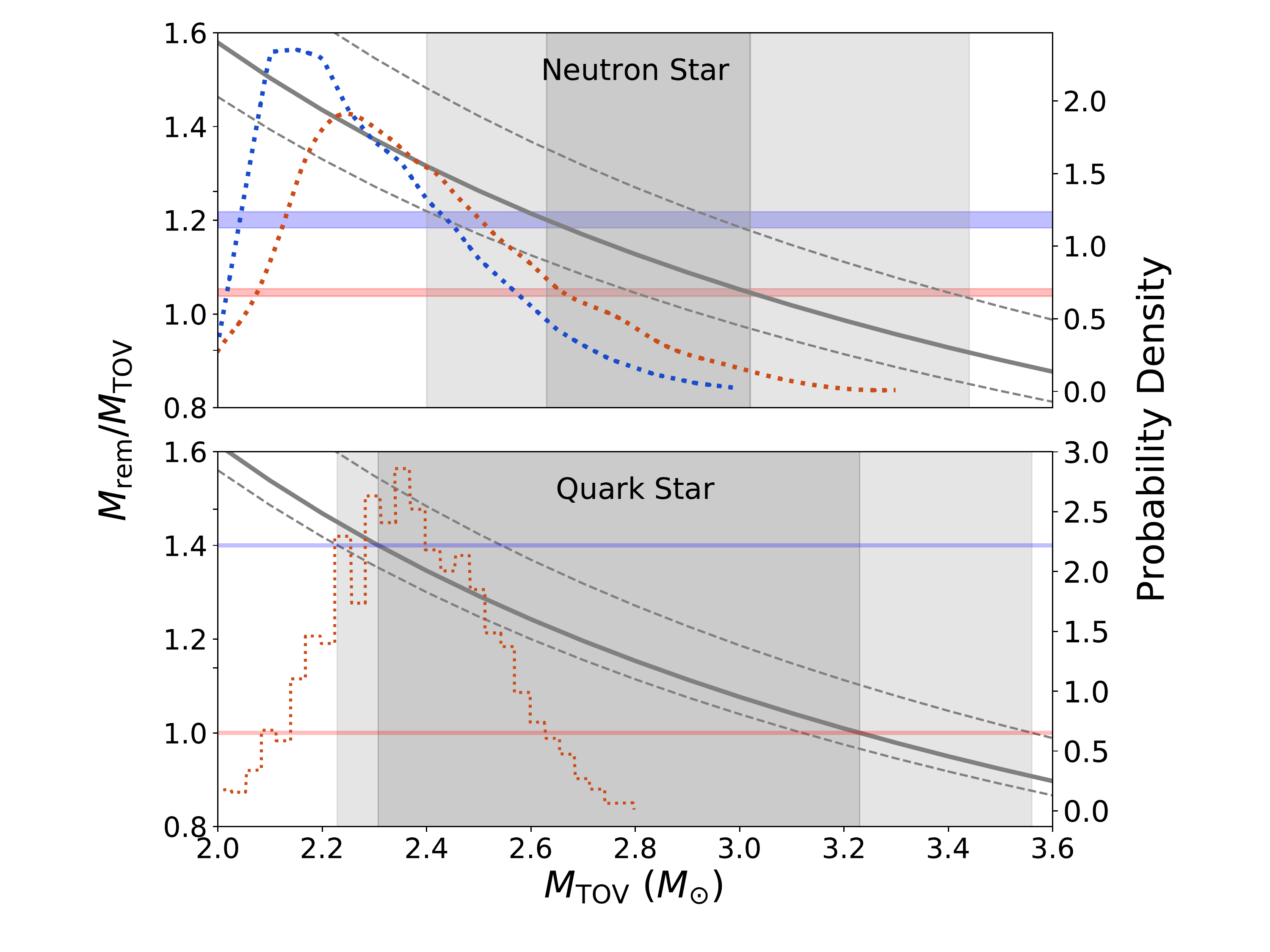}
\caption{{\bf Constraints on the Tolman-Oppenheimer-Volkoff non-spinning maximum mass ($M_{\rm TOV}$) for neutron stars (upper panel) and quark stars (lower panel).} 
Given the derived remnant mass $M_{\rm rem}$ of GW190425, the required $M_{\rm rem}/M_{\rm TOV}$ enhancement factor as a function of $M_{\rm TOV}$ is drawn as the black solid curve with the two associated dashed curves denoting the error range. The maximum enhancement factor can be up to $\sim 1.2$ and $\sim 1.4$ for a uniformly rotating neutron star\cite{cook94,breu16,ai20} and quark star\cite{li16}, respectively. These are shown as the blue solid band in both panels (the width of the band accommodates the possible range of different EoSs). The minimum enhancement factor for a uniformly rotating star is $\sim 1.05$ for neutron stars\cite{ai20}, and we assume $1$ for quark stars as the most conservative limit, which are denoted as red bands in both panels. The allowed range of $M_{\rm TOV}$ given a GW190425/FRB 20190425A association is derived from the intersections of the colored bands and slanted curves; this is shown in dark (light) gray shading when using the central value (including error ranges) of $M_{\rm rem}$.
Published constraints on the distribution of $M_{\rm TOV}$ are presented as colored dotted lines\cite{li21,li21b,miller21} for comparison, with the relevant probability density scale marked on the right axis.}
\label{fig:mtov}
\end{figure}

If confirmed, the astrophysical association between FRB 20190425A and GW190425 can constrain the poorly known equation of state (EoS) of the post-merger compact object. GW parameter estimation reveals that the gravitational masses of the two merger components in the source frame are $M_1 = 2.03^{+0.58}_{-0.34}\ M_{\odot}$ and $M_2 = 1.35^{+0.26}_{-0.26}\ M_{\odot}$\cite{abbott2021}, with the final total
mass of the merger product $M_{\rm tot}^{\rm final} = 3.23^{+0.33}_{-0.11}\ M_{\odot}$ (\url{https://dcc.ligo.org/LIGO-P2000223/public}).
Assuming a uniformly rotating neutron star for the merger remnant, we derive the final gravitational mass $M_{\rm rem} = 3.16^{+0.40}_{-0.24}\ M_{\odot}$ (see Methods), consistent with the final total mass derived from the GW data. There is a universal maximum mass ($M_{\rm TOV})$ that a non-rotating neutron or quark star can support against gravitational collapse into a black hole\cite{breu16,ai20}. This depends on the uncertain neutron star EoS, and is poorly constrained by data\cite{li21,miller21,li21b}. A supramassive neutron star remnant can support a higher mass
with an enhancement factor up to $M_{\rm rem}/M_{\rm TOV} \sim 1.2$ for uniform rotation\cite{cook94,breu16,ai20}. If GW190425 produced a spin-supported supramassive neutron star remnant, this constrains
$M_{\rm TOV} > 2.63^{+0.39}_{-0.23}\ M_{\odot}$. In addition, requiring that the remnant subsequently collapses places an upper limit of $M_{\rm TOV} < M_{\rm rem}/1.046 = 3.02^{+0.42}_{-0.25}\,\mathrm{M_\odot}$ (see Methods). Our constraints on $M_{\rm TOV}$ are at the high end of, but still consistent with, constraints
derived from observations of GW170817 and X-ray emission of Galactic pulsars (Figure~\ref{fig:mtov}, upper panel)
\cite{miller21,li21}. 
Given the high-mass, high-pressure environment of a violent merger, quark deconfinement might occur for the merger remnant\cite{dailu98,drago16}, making the enhancement factor as large as 1.4 (ref.\ \cite{li16}). Assuming $M_\mathrm{rem}= M_{\rm tot}^{\rm final}$, we find $2.31_{-0.08}^{+0.24}\ M_{\odot}<M_{\rm TOV}<3.23^{+0.33}_{-0.11}\ M_{\odot}$, consistent with existing constraints for quark star models\cite{li21b} (Figure~\ref{fig:mtov}, lower panel).

Our constraints on $M_{\rm TOV}$ indicate that mergers with lower component masses such as GW170817 (and its association with GRB 170817A)\cite{abbott2017} would produce a much longer-lived supramassive or stable neutron star. The late-time X-ray rebrightening of the GW170817 remnant is consistent with the existence of such a long-lived remnant \cite{piro19,troja2022}.
These mergers may produce FRBs that repeat over a period of time much longer than 2.5 hours if the magnetar-FRB mechanism applies. This then implies that repeating FRBs could be produced from old stellar populations, such as repeating FRB~20200120E in a globular cluster in M81\cite{CHIME_M81_2021,Kirsten2021_M81GC}. Since the rate density of BNS mergers is much smaller than the inferred rate density of FRBs, this channel alone cannot account for the FRB~20190425A-like FRBs in the CHIME sample. We encourage wide-field radio observations concurrent with future GW observing runs in order to further test our proposed GW--FRB association, and the use of FRB localisation data to identify potential hosts galaxies to guide searches for a multiwavelength counterpart.

\bigskip
\bigskip
\bigskip
\bibliography{nature}

\begin{thebibliography}{10}
\expandafter\ifx\csname url\endcsname\relax
  \def\url#1{\texttt{#1}}\fi
\expandafter\ifx\csname urlprefix\endcsname\relax\def\urlprefix{URL }\fi
\providecommand{\bibinfo}[2]{#2}
\providecommand{\eprint}[2][]{\url{#2}}

\bibitem{lorimer07}
\bibinfo{author}{{Lorimer}, D.~R.}, \bibinfo{author}{{Bailes}, M.},
  \bibinfo{author}{{McLaughlin}, M.~A.}, \bibinfo{author}{{Narkevic}, D.~J.} \&
  \bibinfo{author}{{Crawford}, F.}
\newblock \bibinfo{title}{{A Bright Millisecond Radio Burst of Extragalactic
  Origin}}.
\newblock \emph{\bibinfo{journal}{Science}} \textbf{\bibinfo{volume}{318}},
  \bibinfo{pages}{777} (\bibinfo{year}{2007}).

\bibitem{chime2021}
\bibinfo{author}{Amiri, M.} \emph{et~al.}
\newblock \bibinfo{title}{{The First CHIME/FRB Fast Radio Burst Catalog}}.
\newblock \emph{\bibinfo{journal}{\apjs}} \textbf{\bibinfo{volume}{257}},
  \bibinfo{pages}{59} (\bibinfo{year}{2021}).

\bibitem{Michilli2018_121102}
\bibinfo{author}{{Michilli}, D.} \emph{et~al.}
\newblock \bibinfo{title}{{An extreme magneto-ionic environment associated with
  the fast radio burst source FRB 121102}}.
\newblock \emph{\bibinfo{journal}{\nat}} \textbf{\bibinfo{volume}{553}},
  \bibinfo{pages}{182--185} (\bibinfo{year}{2018}).

\bibitem{chime2020}
\bibinfo{title}{A bright millisecond-duration radio burst from a galactic
  magnetar}.
\newblock \emph{\bibinfo{journal}{Nature}} \textbf{\bibinfo{volume}{587}},
  \bibinfo{pages}{54–58} (\bibinfo{year}{2020}).
\newblock \urlprefix\url{http://dx.doi.org/10.1038/s41586-020-2863-y}.

\bibitem{bochenek2020}
\bibinfo{author}{{Bochenek}, C.~D.} \emph{et~al.}
\newblock \bibinfo{title}{{A fast radio burst associated with a Galactic
  magnetar}}.
\newblock \emph{\bibinfo{journal}{\nat}} \textbf{\bibinfo{volume}{587}},
  \bibinfo{pages}{59--62} (\bibinfo{year}{2020}).

\bibitem{li21c}
\bibinfo{author}{{Li}, C.~K.} \emph{et~al.}
\newblock \bibinfo{title}{{HXMT identification of a non-thermal X-ray burst
  from SGR J1935+2154 and with FRB 200428}}.
\newblock \emph{\bibinfo{journal}{Nature Astronomy}}  (\bibinfo{year}{2021}).

\bibitem{mereghetti20}
\bibinfo{author}{{Mereghetti}, S.} \emph{et~al.}
\newblock \bibinfo{title}{{INTEGRAL Discovery of a Burst with Associated Radio
  Emission from the Magnetar SGR 1935+2154}}.
\newblock \emph{\bibinfo{journal}{\apjl}} \textbf{\bibinfo{volume}{898}},
  \bibinfo{pages}{L29} (\bibinfo{year}{2020}).

\bibitem{Kirsten2021_M81GC}
\bibinfo{author}{{Kirsten}, F.} \emph{et~al.}
\newblock \bibinfo{title}{{A repeating fast radio burst source in a globular
  cluster}}.
\newblock \emph{\bibinfo{journal}{\nat}} \textbf{\bibinfo{volume}{602}},
  \bibinfo{pages}{585--589} (\bibinfo{year}{2022}).

\bibitem{platts2019}
\bibinfo{author}{Platts, E.} \emph{et~al.}
\newblock \bibinfo{title}{{A Living Theory Catalogue for Fast Radio Bursts}}.
\newblock \emph{\bibinfo{journal}{Phys. Rept.}} \textbf{\bibinfo{volume}{821}},
  \bibinfo{pages}{1--27} (\bibinfo{year}{2019}).

\bibitem{aligo2015}
\bibinfo{author}{Aasi, J.} \emph{et~al.}
\newblock \bibinfo{title}{Advanced ligo}.
\newblock \emph{\bibinfo{journal}{Classical and Quantum Gravity}}
  \textbf{\bibinfo{volume}{32}}, \bibinfo{pages}{074001}
  (\bibinfo{year}{2015}).
\newblock \urlprefix\url{http://dx.doi.org/10.1088/0264-9381/32/7/074001}.

\bibitem{acernese2014}
\bibinfo{author}{Acernese, F.} \emph{et~al.}
\newblock \bibinfo{title}{Advanced virgo: a second-generation interferometric
  gravitational wave detector}.
\newblock \emph{\bibinfo{journal}{Classical and Quantum Gravity}}
  \textbf{\bibinfo{volume}{32}}, \bibinfo{pages}{024001}
  (\bibinfo{year}{2014}).
\newblock \urlprefix\url{http://dx.doi.org/10.1088/0264-9381/32/2/024001}.

\bibitem{abbott2020}
\bibinfo{author}{Abbott, B.~P.} \emph{et~al.}
\newblock \bibinfo{title}{Gw190425: Observation of a compact binary coalescence
  with total mass $\sim$ 3.4 m$_{\odot}$}.
\newblock \emph{\bibinfo{journal}{The Astrophysical Journal}}
  \textbf{\bibinfo{volume}{892}}, \bibinfo{pages}{L3} (\bibinfo{year}{2020}).
\newblock \urlprefix\url{http://dx.doi.org/10.3847/2041-8213/ab75f5}.

\bibitem{abbott2021}
\bibinfo{author}{Abbott, R.} \emph{et~al.}
\newblock \bibinfo{title}{Gwtc-2: Compact binary coalescences observed by ligo
  and virgo during the first half of the third observing run}.
\newblock \emph{\bibinfo{journal}{Physical Review X}}
  \textbf{\bibinfo{volume}{11}} (\bibinfo{year}{2021}).
\newblock \urlprefix\url{http://dx.doi.org/10.1103/PhysRevX.11.021053}.

\bibitem{zhang2014}
\bibinfo{author}{Zhang, B.}
\newblock \bibinfo{title}{A possible connection between fast radio bursts and
  gamma-ray bursts}.
\newblock \emph{\bibinfo{journal}{The Astrophysical Journal}}
  \textbf{\bibinfo{volume}{780}}, \bibinfo{pages}{L21} (\bibinfo{year}{2013}).
\newblock \urlprefix\url{http://dx.doi.org/10.1088/2041-8205/780/2/L21}.

\bibitem{falcke2014}
\bibinfo{author}{{Falcke}, H.} \& \bibinfo{author}{{Rezzolla}, L.}
\newblock \bibinfo{title}{{Fast radio bursts: the last sign of supramassive
  neutron stars}}.
\newblock \emph{\bibinfo{journal}{\aap}} \textbf{\bibinfo{volume}{562}},
  \bibinfo{pages}{A137} (\bibinfo{year}{2014}).

\bibitem{tov1939}
\bibinfo{author}{Oppenheimer, J.~R.} \& \bibinfo{author}{Volkoff, G.~M.}
\newblock \bibinfo{title}{On massive neutron cores}.
\newblock \emph{\bibinfo{journal}{Phys. Rev.}} \textbf{\bibinfo{volume}{55}},
  \bibinfo{pages}{374--381} (\bibinfo{year}{1939}).
\newblock \urlprefix\url{https://link.aps.org/doi/10.1103/PhysRev.55.374}.

\bibitem{pleunis2021}
\bibinfo{author}{Pleunis, Z.} \emph{et~al.}
\newblock \bibinfo{title}{Lofar detection of 110–188 mhz emission and
  frequency-dependent activity from frb 20180916b}.
\newblock \emph{\bibinfo{journal}{The Astrophysical Journal Letters}}
  \textbf{\bibinfo{volume}{911}} (\bibinfo{year}{2020}).

\bibitem{gajjar2018}
\bibinfo{author}{Gajjar, V.} \emph{et~al.}
\newblock \bibinfo{title}{Highest frequency detection of frb 121102 at 4–8
  ghz using the breakthrough listen digital backend at the green bank
  telescope}.
\newblock \emph{\bibinfo{journal}{The Astrophysical Journal}}
  \textbf{\bibinfo{volume}{863}}, \bibinfo{pages}{2} (\bibinfo{year}{2018}).
\newblock \urlprefix\url{http://dx.doi.org/10.3847/1538-4357/aad005}.

\bibitem{ravi2019}
\bibinfo{author}{Ravi, V.}
\newblock \bibinfo{title}{The prevalence of repeating fast radio bursts}.
\newblock \emph{\bibinfo{journal}{Nature Astronomy}}
  \textbf{\bibinfo{volume}{3}}, \bibinfo{pages}{928–931}
  (\bibinfo{year}{2019}).
\newblock \urlprefix\url{http://dx.doi.org/10.1038/s41550-019-0831-y}.

\bibitem{luo2020}
\bibinfo{author}{Luo, R.} \emph{et~al.}
\newblock \bibinfo{title}{On the frb luminosity function – – ii. event rate
  density}.
\newblock \emph{\bibinfo{journal}{Monthly Notices of the Royal Astronomical
  Society}} \textbf{\bibinfo{volume}{494}}, \bibinfo{pages}{665–679}
  (\bibinfo{year}{2020}).
\newblock \urlprefix\url{http://dx.doi.org/10.1093/mnras/staa704}.

\bibitem{spitler2016}
\bibinfo{author}{Spitler, L.~G.} \emph{et~al.}
\newblock \bibinfo{title}{A repeating fast radio burst}.
\newblock \emph{\bibinfo{journal}{Nature}} \textbf{\bibinfo{volume}{531}},
  \bibinfo{pages}{202–205} (\bibinfo{year}{2016}).
\newblock \urlprefix\url{http://dx.doi.org/10.1038/nature17168}.

\bibitem{James2020a}
\bibinfo{author}{{James}, C.~W.} \emph{et~al.}
\newblock \bibinfo{title}{{Which bright fast radio bursts repeat?}}
\newblock \emph{\bibinfo{journal}{\mnras}} \textbf{\bibinfo{volume}{495}},
  \bibinfo{pages}{2416--2427} (\bibinfo{year}{2020}).

\bibitem{piro2012}
\bibinfo{author}{Piro, A.~L.}
\newblock \bibinfo{title}{Magnetic interactions in coalescing neutron star
  binaries}.
\newblock \emph{\bibinfo{journal}{The Astrophysical Journal}}
  \textbf{\bibinfo{volume}{755}}, \bibinfo{pages}{80} (\bibinfo{year}{2012}).
\newblock \urlprefix\url{http://dx.doi.org/10.1088/0004-637X/755/1/80}.

\bibitem{zhang2020b}
\bibinfo{author}{Zhang, B.}
\newblock \bibinfo{title}{Fast radio bursts from interacting binary neutron
  star systems}.
\newblock \emph{\bibinfo{journal}{The Astrophysical Journal}}
  \textbf{\bibinfo{volume}{890}}, \bibinfo{pages}{L24} (\bibinfo{year}{2020}).
\newblock \urlprefix\url{http://dx.doi.org/10.3847/2041-8213/ab7244}.

\bibitem{totani2013}
\bibinfo{author}{Totani, T.}
\newblock \bibinfo{title}{Cosmological fast radio bursts from binary neutron
  star mergers}.
\newblock \emph{\bibinfo{journal}{Publications of the Astronomical Society of
  Japan}} \textbf{\bibinfo{volume}{65}}, \bibinfo{pages}{L12}
  (\bibinfo{year}{2013}).
\newblock \urlprefix\url{http://dx.doi.org/10.1093/pasj/65.5.L12}.

\bibitem{abbott2020b}
\bibinfo{author}{Abbott, B.~P.} \emph{et~al.}
\newblock \bibinfo{title}{Prospects for observing and localizing
  gravitational-wave transients with advanced ligo, advanced virgo and kagra}.
\newblock \emph{\bibinfo{journal}{Living Reviews in Relativity}}
  \textbf{\bibinfo{volume}{23}} (\bibinfo{year}{2020}).
\newblock \urlprefix\url{http://dx.doi.org/10.1007/s41114-020-00026-9}.

\bibitem{macquart2020}
\bibinfo{author}{{Macquart}, J.~P.} \emph{et~al.}
\newblock \bibinfo{title}{{A census of baryons in the Universe from localized
  fast radio bursts}}.
\newblock \emph{\bibinfo{journal}{\nat}} \textbf{\bibinfo{volume}{581}},
  \bibinfo{pages}{391--395} (\bibinfo{year}{2020}).

\bibitem{CHIME2021_lat_dependence}
\bibinfo{author}{{Josephy}, A.} \emph{et~al.}
\newblock \bibinfo{title}{{No Evidence for Galactic Latitude Dependence of the
  Fast Radio Burst Sky Distribution}}.
\newblock \emph{\bibinfo{journal}{arXiv e-prints}}
  \bibinfo{pages}{arXiv:2106.04353} (\bibinfo{year}{2021}).

\bibitem{CHIME2021_structure}
\bibinfo{author}{{Rafiei-Ravandi}, M.} \emph{et~al.}
\newblock \bibinfo{title}{{CHIME/FRB Catalog 1 results: statistical
  cross-correlations with large-scale structure}}.
\newblock \emph{\bibinfo{journal}{arXiv e-prints}}
  \bibinfo{pages}{arXiv:2106.04354} (\bibinfo{year}{2021}).

\bibitem{Gehrels2016}
\bibinfo{author}{{Gehrels}, N.} \emph{et~al.}
\newblock \bibinfo{title}{{Galaxy Strategy for LIGO-Virgo Gravitational Wave
  Counterpart Searches}}.
\newblock \emph{\bibinfo{journal}{\apj}} \textbf{\bibinfo{volume}{820}},
  \bibinfo{pages}{136} (\bibinfo{year}{2016}).

\bibitem{nakar11}
\bibinfo{author}{{Nakar}, E.} \& \bibinfo{author}{{Piran}, T.}
\newblock \bibinfo{title}{{Detectable radio flares following gravitational
  waves from mergers of binary neutron stars}}.
\newblock \emph{\bibinfo{journal}{\nat}} \textbf{\bibinfo{volume}{478}},
  \bibinfo{pages}{82--84} (\bibinfo{year}{2011}).

\bibitem{gao13}
\bibinfo{author}{{Gao}, H.}, \bibinfo{author}{{Ding}, X.},
  \bibinfo{author}{{Wu}, X.-F.}, \bibinfo{author}{{Zhang}, B.} \&
  \bibinfo{author}{{Dai}, Z.-G.}
\newblock \bibinfo{title}{{Bright Broadband Afterglows of Gravitational Wave
  Bursts from Mergers of Binary Neutron Stars}}.
\newblock \emph{\bibinfo{journal}{\apj}} \textbf{\bibinfo{volume}{771}},
  \bibinfo{pages}{86} (\bibinfo{year}{2013}).

\bibitem{most2018}
\bibinfo{author}{{Most}, E.~R.}, \bibinfo{author}{{Nathanail}, A.} \&
  \bibinfo{author}{{Rezzolla}, L.}
\newblock \bibinfo{title}{{Electromagnetic Emission from Blitzars and Its
  Impact on Non-repeating Fast Radio Bursts}}.
\newblock \emph{\bibinfo{journal}{\apj}} \textbf{\bibinfo{volume}{864}},
  \bibinfo{pages}{117} (\bibinfo{year}{2018}).

\bibitem{abbott2017b}
\bibinfo{author}{Abbott, B.~P.} \emph{et~al.}
\newblock \bibinfo{title}{Gravitational waves and gamma-rays from a binary
  neutron star merger: Gw170817 and grb 170817a}.
\newblock \emph{\bibinfo{journal}{The Astrophysical Journal}}
  \textbf{\bibinfo{volume}{848}}, \bibinfo{pages}{L13} (\bibinfo{year}{2017}).
\newblock \urlprefix\url{http://dx.doi.org/10.3847/2041-8213/aa920c}.

\bibitem{zhangbb2018}
\bibinfo{author}{{Zhang}, B.~B.} \emph{et~al.}
\newblock \bibinfo{title}{{A peculiar low-luminosity short gamma-ray burst from
  a double neutron star merger progenitor}}.
\newblock \emph{\bibinfo{journal}{Nature Communications}}
  \textbf{\bibinfo{volume}{9}}, \bibinfo{pages}{447} (\bibinfo{year}{2018}).

\bibitem{INTEGRALGCN2}
\bibinfo{author}{{Savchenko}, V.} \emph{et~al.}
\newblock \bibinfo{title}{{LIGO/Virgo S190425z: further analysis of INTEGRAL
  data.}}
\newblock \emph{\bibinfo{journal}{GRB Coordinates Network}}
  \textbf{\bibinfo{volume}{24178}}, \bibinfo{pages}{1} (\bibinfo{year}{2019}).

\bibitem{pozanenko2019}
\bibinfo{author}{Pozanenko, A.~S.}, \bibinfo{author}{Minaev, P.~Y.},
  \bibinfo{author}{Grebenev, S.~A.} \& \bibinfo{author}{Chelovekov, I.~V.}
\newblock \bibinfo{title}{Observation of the second ligo/virgo event connected
  with a binary neutron star merger s190425z in the gamma-ray range}.
\newblock \emph{\bibinfo{journal}{Astronomy Letters}}
  \textbf{\bibinfo{volume}{45}}, \bibinfo{pages}{710–727}
  (\bibinfo{year}{2019}).
\newblock \urlprefix\url{http://dx.doi.org/10.1134/S1063773719110057}.

\bibitem{breu16}
\bibinfo{author}{{Breu}, C.} \& \bibinfo{author}{{Rezzolla}, L.}
\newblock \bibinfo{title}{{Maximum mass, moment of inertia and compactness of
  relativistic stars}}.
\newblock \emph{\bibinfo{journal}{\mnras}} \textbf{\bibinfo{volume}{459}},
  \bibinfo{pages}{646--656} (\bibinfo{year}{2016}).

\bibitem{ai20}
\bibinfo{author}{{Ai}, S.}, \bibinfo{author}{{Gao}, H.} \&
  \bibinfo{author}{{Zhang}, B.}
\newblock \bibinfo{title}{{What Constraints on the Neutron Star Maximum Mass
  Can One Pose from GW170817 Observations?}}
\newblock \emph{\bibinfo{journal}{\apj}} \textbf{\bibinfo{volume}{893}},
  \bibinfo{pages}{146} (\bibinfo{year}{2020}).

\bibitem{li21}
\bibinfo{author}{{Li}, A.}, \bibinfo{author}{{Miao}, Z.},
  \bibinfo{author}{{Han}, S.} \& \bibinfo{author}{{Zhang}, B.}
\newblock \bibinfo{title}{{Constraints on the Maximum Mass of Neutron Stars
  with a Quark Core from GW170817 and NICER PSR J0030+0451 Data}}.
\newblock \emph{\bibinfo{journal}{\apj}} \textbf{\bibinfo{volume}{913}},
  \bibinfo{pages}{27} (\bibinfo{year}{2021}).

\bibitem{miller21}
\bibinfo{author}{{Miller}, M.~C.} \emph{et~al.}
\newblock \bibinfo{title}{{The Radius of PSR J0740+6620 from NICER and
  XMM-Newton Data}}.
\newblock \emph{\bibinfo{journal}{arXiv e-prints}}
  \bibinfo{pages}{arXiv:2105.06979} (\bibinfo{year}{2021}).

\bibitem{li21b}
\bibinfo{author}{{Li}, A.}, \bibinfo{author}{{Miao}, Z.~Q.},
  \bibinfo{author}{{Jiang}, J.~L.}, \bibinfo{author}{{Tang}, S.~P.} \&
  \bibinfo{author}{{Xu}, R.~X.}
\newblock \bibinfo{title}{{Bayesian inference of quark star equation of state
  using the NICER PSR J0030+0451 data}}.
\newblock \emph{\bibinfo{journal}{\mnras}}  (\bibinfo{year}{2021}).

\bibitem{cook94}
\bibinfo{author}{{Cook}, G.~B.}, \bibinfo{author}{{Shapiro}, S.~L.} \&
  \bibinfo{author}{{Teukolsky}, S.~A.}
\newblock \bibinfo{title}{{Rapidly Rotating Neutron Stars in General
  Relativity: Realistic Equations of State}}.
\newblock \emph{\bibinfo{journal}{\apj}} \textbf{\bibinfo{volume}{424}},
  \bibinfo{pages}{823} (\bibinfo{year}{1994}).

\bibitem{dailu98}
\bibinfo{author}{{Dai}, Z.~G.} \& \bibinfo{author}{{Lu}, T.}
\newblock \bibinfo{title}{{{\ensuremath{\gamma}}-Ray Bursts and Afterglows from
  Rotating Strange Stars and Neutron Stars}}.
\newblock \emph{\bibinfo{journal}{\prl}} \textbf{\bibinfo{volume}{81}},
  \bibinfo{pages}{4301--4304} (\bibinfo{year}{1998}).

\bibitem{drago16}
\bibinfo{author}{{Drago}, A.}, \bibinfo{author}{{Lavagno}, A.},
  \bibinfo{author}{{Metzger}, B.~D.} \& \bibinfo{author}{{Pagliara}, G.}
\newblock \bibinfo{title}{{Quark deconfinement and the duration of short
  gamma-ray bursts}}.
\newblock \emph{\bibinfo{journal}{\prd}} \textbf{\bibinfo{volume}{93}},
  \bibinfo{pages}{103001} (\bibinfo{year}{2016}).

\bibitem{li16}
\bibinfo{author}{{Li}, A.} \emph{et~al.}
\newblock \bibinfo{title}{{Internal x-ray plateau in short GRBs: Signature of
  supramassive fast-rotating quark stars?}}
\newblock \emph{\bibinfo{journal}{\prd}} \textbf{\bibinfo{volume}{94}},
  \bibinfo{pages}{083010} (\bibinfo{year}{2016}).

\bibitem{abbott2017}
\bibinfo{author}{Abbott, B.~P.} \emph{et~al.}
\newblock \bibinfo{title}{Gw170817: Observation of gravitational waves from a
  binary neutron star inspiral}.
\newblock \emph{\bibinfo{journal}{Physical Review Letters}}
  \textbf{\bibinfo{volume}{119}} (\bibinfo{year}{2017}).
\newblock \urlprefix\url{http://dx.doi.org/10.1103/PhysRevLett.119.161101}.

\bibitem{piro19}
\bibinfo{author}{{Piro}, L.} \emph{et~al.}
\newblock \bibinfo{title}{{A long-lived neutron star merger remnant in
  GW170817: constraints and clues from X-ray observations}}.
\newblock \emph{\bibinfo{journal}{\mnras}} \textbf{\bibinfo{volume}{483}},
  \bibinfo{pages}{1912--1921} (\bibinfo{year}{2019}).

\bibitem{troja2022}
\bibinfo{author}{{Troja}, E.} \emph{et~al.}
\newblock \bibinfo{title}{{Accurate flux calibration of GW170817: is the X-ray
  counterpart on the rise?}}
\newblock \emph{\bibinfo{journal}{\mnras}} \textbf{\bibinfo{volume}{510}},
  \bibinfo{pages}{1902--1909} (\bibinfo{year}{2022}).

\bibitem{CHIME_M81_2021}
\bibinfo{author}{{Bhardwaj}, M.} \emph{et~al.}
\newblock \bibinfo{title}{{A Nearby Repeating Fast Radio Burst in the Direction
  of M81}}.
\newblock \emph{\bibinfo{journal}{\apjl}} \textbf{\bibinfo{volume}{910}},
  \bibinfo{pages}{L18} (\bibinfo{year}{2021}).

\end{thebibliography}


\begin{thebibliography}{10}
\expandafter\ifx\csname url\endcsname\relax
  \def\url#1{\texttt{#1}}\fi
\expandafter\ifx\csname urlprefix\endcsname\relax\def\urlprefix{URL }\fi
\providecommand{\bibinfo}[2]{#2}
\providecommand{\eprint}[2][]{\url{#2}}

\bibitem{zhang2020}
\bibinfo{author}{Zhang, G.~Q.}, \bibinfo{author}{Yi, S.~X.} \&
  \bibinfo{author}{Wang, F.~Y.}
\newblock \bibinfo{title}{The rarity of repeating fast radio bursts from binary
  neutron star mergers}.
\newblock \emph{\bibinfo{journal}{The Astrophysical Journal}}
  \textbf{\bibinfo{volume}{893}}, \bibinfo{pages}{44} (\bibinfo{year}{2020}).
\newblock \urlprefix\url{http://dx.doi.org/10.3847/1538-4357/ab7c5c}.

\bibitem{Alex2021thesis}
\bibinfo{author}{{Moroianu}, A.}
\newblock \emph{\bibinfo{title}{Deep Search for Fast Radio Burst and
  Gravitational Wave Coincidences}}.
\newblock \bibinfo{type}{Masters thesis}, \bibinfo{school}{University of
  Western Australia} (\bibinfo{year}{2021}).

\bibitem{piro2012}
\bibinfo{author}{Piro, A.~L.}
\newblock \bibinfo{title}{Magnetic interactions in coalescing neutron star
  binaries}.
\newblock \emph{\bibinfo{journal}{The Astrophysical Journal}}
  \textbf{\bibinfo{volume}{755}}, \bibinfo{pages}{80} (\bibinfo{year}{2012}).
\newblock \urlprefix\url{http://dx.doi.org/10.1088/0004-637X/755/1/80}.

\bibitem{wang2016}
\bibinfo{author}{Wang, J.-S.}, \bibinfo{author}{Yang, Y.-P.},
  \bibinfo{author}{Wu, X.-F.}, \bibinfo{author}{Dai, Z.-G.} \&
  \bibinfo{author}{Wang, F.-Y.}
\newblock \bibinfo{title}{Fast radio bursts from the inspiral of double neutron
  stars}.
\newblock \emph{\bibinfo{journal}{The Astrophysical Journal}}
  \textbf{\bibinfo{volume}{822}}, \bibinfo{pages}{L7} (\bibinfo{year}{2016}).
\newblock \urlprefix\url{http://dx.doi.org/10.3847/2041-8205/822/1/L7}.

\bibitem{zhang2016}
\bibinfo{author}{Zhang, B.}
\newblock \bibinfo{title}{Mergers of charged black holes: Gravitational-wave
  events, short gamma-ray bursts, and fast radio bursts}.
\newblock \emph{\bibinfo{journal}{The Astrophysical Journal}}
  \textbf{\bibinfo{volume}{827}}, \bibinfo{pages}{L31} (\bibinfo{year}{2016}).
\newblock \urlprefix\url{http://dx.doi.org/10.3847/2041-8205/827/2/L31}.

\bibitem{totani2013}
\bibinfo{author}{Totani, T.}
\newblock \bibinfo{title}{Cosmological fast radio bursts from binary neutron
  star mergers}.
\newblock \emph{\bibinfo{journal}{Publications of the Astronomical Society of
  Japan}} \textbf{\bibinfo{volume}{65}}, \bibinfo{pages}{L12}
  (\bibinfo{year}{2013}).
\newblock \urlprefix\url{http://dx.doi.org/10.1093/pasj/65.5.L12}.

\bibitem{zhang2014}
\bibinfo{author}{Zhang, B.}
\newblock \bibinfo{title}{A possible connection between fast radio bursts and
  gamma-ray bursts}.
\newblock \emph{\bibinfo{journal}{The Astrophysical Journal}}
  \textbf{\bibinfo{volume}{780}}, \bibinfo{pages}{L21} (\bibinfo{year}{2013}).
\newblock \urlprefix\url{http://dx.doi.org/10.1088/2041-8205/780/2/L21}.

\bibitem{wen2010}
\bibinfo{author}{Wen, L.} \& \bibinfo{author}{Chen, Y.}
\newblock \bibinfo{title}{Geometrical expression for the angular resolution of
  a network of gravitational-wave detectors}.
\newblock \emph{\bibinfo{journal}{Physical Review D}}
  \textbf{\bibinfo{volume}{81}} (\bibinfo{year}{2010}).
\newblock \urlprefix\url{http://dx.doi.org/10.1103/PhysRevD.81.082001}.

\bibitem{fairhurst2011}
\bibinfo{author}{Fairhurst, S.}
\newblock \bibinfo{title}{Triangulation of gravitational wave sources with a
  network of detectors}.
\newblock \emph{\bibinfo{journal}{New Journal of Physics}}
  \textbf{\bibinfo{volume}{13}}, \bibinfo{pages}{069602}
  (\bibinfo{year}{2011}).
\newblock \urlprefix\url{https://doi.org/10.1088/1367-2630/13/6/069602}.

\bibitem{klimenko2011}
\bibinfo{author}{Klimenko, S.} \emph{et~al.}
\newblock \bibinfo{title}{Localization of gravitational wave sources with
  networks of advanced detectors}.
\newblock \emph{\bibinfo{journal}{Physical Review D}}
  \textbf{\bibinfo{volume}{83}} (\bibinfo{year}{2011}).

\bibitem{pankow2018}
\bibinfo{author}{Pankow, C.}, \bibinfo{author}{Chase, E.~A.},
  \bibinfo{author}{Coughlin, S.}, \bibinfo{author}{Zevin, M.} \&
  \bibinfo{author}{Kalogera, V.}
\newblock \bibinfo{title}{Improvements in gravitational-wave sky localization
  with expanded networks of interferometers}.
\newblock \emph{\bibinfo{journal}{The Astrophysical Journal}}
  \textbf{\bibinfo{volume}{854}}, \bibinfo{pages}{L25} (\bibinfo{year}{2018}).
\newblock \urlprefix\url{http://dx.doi.org/10.3847/2041-8213/aaacd4}.

\bibitem{singer2016}
\bibinfo{author}{Singer, L.~P.} \& \bibinfo{author}{Price, L.~R.}
\newblock \bibinfo{title}{Rapid bayesian position reconstruction for
  gravitational-wave transients}.
\newblock \emph{\bibinfo{journal}{Physical Review D}}
  \textbf{\bibinfo{volume}{93}} (\bibinfo{year}{2016}).
\newblock \urlprefix\url{http://dx.doi.org/10.1103/PhysRevD.93.024013}.

\bibitem{sivia2006}
\bibinfo{author}{Sivia, D.~S.} \& \bibinfo{author}{Skilling, J.}
\newblock \emph{\bibinfo{title}{{Data Analysis - A Bayesian Tutorial}}}.
\newblock Oxford Science Publications (\bibinfo{publisher}{Oxford University
  Press}, \bibinfo{year}{2006}), \bibinfo{edition}{2nd} edn.

\bibitem{abbott2020b}
\bibinfo{author}{Abbott, B.~P.} \emph{et~al.}
\newblock \bibinfo{title}{Prospects for observing and localizing
  gravitational-wave transients with advanced ligo, advanced virgo and kagra}.
\newblock \emph{\bibinfo{journal}{Living Reviews in Relativity}}
  \textbf{\bibinfo{volume}{23}} (\bibinfo{year}{2020}).
\newblock \urlprefix\url{http://dx.doi.org/10.1007/s41114-020-00026-9}.

\bibitem{ligoskymap}
\bibinfo{author}{Singer, L.}
\newblock \bibinfo{title}{ligo.skymap}.
\newblock \urlprefix\url{https://git.ligo.org/lscsoft/ligo.skymap}.

\bibitem{macquart2020}
\bibinfo{author}{{Macquart}, J.~P.} \emph{et~al.}
\newblock \bibinfo{title}{{A census of baryons in the Universe from localized
  fast radio bursts}}.
\newblock \emph{\bibinfo{journal}{\nat}} \textbf{\bibinfo{volume}{581}},
  \bibinfo{pages}{391--395} (\bibinfo{year}{2020}).

\bibitem{CHIME2020_periodic}
\bibinfo{author}{{Chime/Frb Collaboration}}, \bibinfo{author}{{Amiri}, M.}
  \emph{et~al.}
\newblock \bibinfo{title}{{Periodic activity from a fast radio burst source}}.
\newblock \emph{\bibinfo{journal}{\nat}} \textbf{\bibinfo{volume}{582}},
  \bibinfo{pages}{351--355} (\bibinfo{year}{2020}).

\bibitem{Rajwade_121102_repetitions}
\bibinfo{author}{{Rajwade}, K.~M.} \emph{et~al.}
\newblock \bibinfo{title}{{Possible periodic activity in the repeating FRB
  121102}}.
\newblock \emph{\bibinfo{journal}{\mnras}} \textbf{\bibinfo{volume}{495}},
  \bibinfo{pages}{3551--3558} (\bibinfo{year}{2020}).

\bibitem{chime2020}
\bibinfo{author}{{CHIME/FRB Collaboration}} \emph{et~al.}
\newblock \bibinfo{title}{{A bright millisecond-duration radio burst from a
  Galactic magnetar}}.
\newblock \emph{\bibinfo{journal}{\nat}} \textbf{\bibinfo{volume}{587}},
  \bibinfo{pages}{54--58} (\bibinfo{year}{2020}).

\bibitem{CHIME2021_lat_dependence}
\bibinfo{author}{{Josephy}, A.} \emph{et~al.}
\newblock \bibinfo{title}{{No Evidence for Galactic Latitude Dependence of the
  Fast Radio Burst Sky Distribution}}.
\newblock \emph{\bibinfo{journal}{arXiv e-prints}}
  \bibinfo{pages}{arXiv:2106.04353} (\bibinfo{year}{2021}).

\bibitem{CHIME2021_structure}
\bibinfo{author}{{Rafiei-Ravandi}, M.} \emph{et~al.}
\newblock \bibinfo{title}{{CHIME/FRB Catalog 1 results: statistical
  cross-correlations with large-scale structure}}.
\newblock \emph{\bibinfo{journal}{arXiv e-prints}}
  \bibinfo{pages}{arXiv:2106.04354} (\bibinfo{year}{2021}).

\bibitem{platts2019}
\bibinfo{author}{Platts, E.} \emph{et~al.}
\newblock \bibinfo{title}{{A Living Theory Catalogue for Fast Radio Bursts}}.
\newblock \emph{\bibinfo{journal}{Phys. Rept.}} \textbf{\bibinfo{volume}{821}},
  \bibinfo{pages}{1--27} (\bibinfo{year}{2019}).

\bibitem{LVKchime2022}
\bibinfo{author}{{The LIGO Scientific Collaboration}} \emph{et~al.}
\newblock \bibinfo{title}{{Search for Gravitational Waves Associated with Fast
  Radio Bursts Detected by CHIME/FRB During the LIGO--Virgo Observing Run
  O3a}}.
\newblock \emph{\bibinfo{journal}{arXiv e-prints}}
  \bibinfo{pages}{arXiv:2203.12038} (\bibinfo{year}{2022}).

\bibitem{Zhang2016bbh}
\bibinfo{author}{{Zhang}, B.}
\newblock \bibinfo{title}{{Mergers of Charged Black Holes: Gravitational-wave
  Events, Short Gamma-Ray Bursts, and Fast Radio Bursts}}.
\newblock \emph{\bibinfo{journal}{\apjl}} \textbf{\bibinfo{volume}{827}},
  \bibinfo{pages}{L31} (\bibinfo{year}{2016}).

\bibitem{lalinference}
\bibinfo{author}{Veitch, J.} \emph{et~al.}
\newblock \bibinfo{title}{Parameter estimation for compact binaries with
  ground-based gravitational-wave observations using the lalinference software
  library}.
\newblock \emph{\bibinfo{journal}{Phys. Rev. D}} \textbf{\bibinfo{volume}{91}},
  \bibinfo{pages}{042003} (\bibinfo{year}{2015}).
\newblock \urlprefix\url{https://link.aps.org/doi/10.1103/PhysRevD.91.042003}.

\bibitem{cordes2002}
\bibinfo{author}{{Cordes}, J.~M.} \& \bibinfo{author}{{Lazio}, T.~J.~W.}
\newblock \bibinfo{title}{{NE2001.I. A New Model for the Galactic Distribution
  of Free Electrons and its Fluctuations}}.
\newblock \emph{\bibinfo{journal}{arXiv e-prints}}
  \bibinfo{pages}{astro--ph/0207156} (\bibinfo{year}{2002}).

\bibitem{yao2017}
\bibinfo{author}{Yao, J.~M.}, \bibinfo{author}{Manchester, R.~N.} \&
  \bibinfo{author}{Wang, N.}
\newblock \bibinfo{title}{A new electron-density model for estimation of pulsar
  and frb distances}.
\newblock \emph{\bibinfo{journal}{The Astrophysical Journal}}
  \textbf{\bibinfo{volume}{835}}, \bibinfo{pages}{29} (\bibinfo{year}{2017}).
\newblock \urlprefix\url{http://dx.doi.org/10.3847/1538-4357/835/1/29}.

\bibitem{yamasaki2020}
\bibinfo{author}{Yamasaki, S.} \& \bibinfo{author}{Totani, T.}
\newblock \bibinfo{title}{The galactic halo contribution to the dispersion
  measure of extragalactic fast radio bursts}.
\newblock \emph{\bibinfo{journal}{The Astrophysical Journal}}
  \textbf{\bibinfo{volume}{888}}, \bibinfo{pages}{105} (\bibinfo{year}{2020}).
\newblock \urlprefix\url{http://dx.doi.org/10.3847/1538-4357/ab58c4}.

\bibitem{Schnitzeler2012DM}
\bibinfo{author}{{Schnitzeler}, D.~H.~F.~M.}
\newblock \bibinfo{title}{{Modelling the Galactic distribution of free
  electrons}}.
\newblock \emph{\bibinfo{journal}{\mnras}} \textbf{\bibinfo{volume}{427}},
  \bibinfo{pages}{664--678} (\bibinfo{year}{2012}).

\bibitem{Shannon2018}
\bibinfo{author}{{Shannon}, R.~M.} \emph{et~al.}
\newblock \bibinfo{title}{{The dispersion-brightness relation for fast radio
  bursts from a wide-field survey.}}
\newblock \emph{\bibinfo{journal}{\nat}} \textbf{\bibinfo{volume}{562}},
  \bibinfo{pages}{386--390} (\bibinfo{year}{2018}).

\bibitem{CHIME_M81_2021}
\bibinfo{author}{{Bhardwaj}, M.} \emph{et~al.}
\newblock \bibinfo{title}{{A Nearby Repeating Fast Radio Burst in the Direction
  of M81}}.
\newblock \emph{\bibinfo{journal}{\apjl}} \textbf{\bibinfo{volume}{910}},
  \bibinfo{pages}{L18} (\bibinfo{year}{2021}).

\bibitem{ProchaskaZheng2019}
\bibinfo{author}{{Prochaska}, J.~X.} \& \bibinfo{author}{{Zheng}, Y.}
\newblock \bibinfo{title}{{Probing Galactic haloes with fast radio bursts}}.
\newblock \emph{\bibinfo{journal}{\mnras}} \textbf{\bibinfo{volume}{485}},
  \bibinfo{pages}{648--665} (\bibinfo{year}{2019}).

\bibitem{Inoue2004}
\bibinfo{author}{{Inoue}, S.}
\newblock \bibinfo{title}{{Probing the cosmic reionization history and local
  environment of gamma-ray bursts through radio dispersion}}.
\newblock \emph{\bibinfo{journal}{\mnras}} \textbf{\bibinfo{volume}{348}},
  \bibinfo{pages}{999--1008} (\bibinfo{year}{2004}).

\bibitem{ade2016}
\bibinfo{author}{Ade, P. A.~R.} \emph{et~al.}
\newblock \bibinfo{title}{Planck2015 results}.
\newblock \emph{\bibinfo{journal}{Astronomy \& Astrophysics}}
  \textbf{\bibinfo{volume}{594}}, \bibinfo{pages}{A13} (\bibinfo{year}{2016}).
\newblock \urlprefix\url{http://dx.doi.org/10.1051/0004-6361/201525830}.

\bibitem{Wright_2006_cosmology}
\bibinfo{author}{{Wright}, E.~L.}
\newblock \bibinfo{title}{{A Cosmology Calculator for the World Wide Web}}.
\newblock \emph{\bibinfo{journal}{\pasp}} \textbf{\bibinfo{volume}{118}},
  \bibinfo{pages}{1711--1715} (\bibinfo{year}{2006}).

\bibitem{James2021}
\bibinfo{author}{{James}, C.~W.} \emph{et~al.}
\newblock \bibinfo{title}{{The z-DM distribution of fast radio bursts}}.
\newblock \emph{\bibinfo{journal}{\mnras}} \textbf{\bibinfo{volume}{509}},
  \bibinfo{pages}{4775--4802} (\bibinfo{year}{2022}).

\bibitem{frb}
\bibinfo{author}{Neeleman, J. X. P. S. S. C. L. N. T.~M.}
\newblock \bibinfo{title}{Frb}.
\newblock \urlprefix\url{https://zenodo.org/record/3403651#.YRxkcBMzZKA}.

\bibitem{zdm}
\bibinfo{author}{Clancy W.~James, E. M.~G., J. Xavier~Prochaska}.
\newblock \bibinfo{title}{zdm}.
\newblock \urlprefix\url{https://zenodo.org/record/5213780#.YRxh5BMzZKA}.

\bibitem{PlanckCosmology2018}
\bibinfo{author}{Aghanim, N.} \emph{et~al.}
\newblock \bibinfo{title}{Planck 2018 results}.
\newblock \emph{\bibinfo{journal}{Astronomy \& Astrophysics}}
  \textbf{\bibinfo{volume}{641}}, \bibinfo{pages}{A6} (\bibinfo{year}{2020}).
\newblock \urlprefix\url{http://dx.doi.org/10.1051/0004-6361/201833910}.

\bibitem{ashton2018}
\bibinfo{author}{{Ashton}, G.} \emph{et~al.}
\newblock \bibinfo{title}{{Coincident Detection Significance in Multimessenger
  Astronomy}}.
\newblock \emph{\bibinfo{journal}{\apj}} \textbf{\bibinfo{volume}{860}},
  \bibinfo{pages}{6} (\bibinfo{year}{2018}).

\bibitem{Nitz2021}
\bibinfo{author}{Nitz, A.} \emph{et~al.}
\newblock \bibinfo{title}{gwastro/pycbc: Release v1.18.3 of pycbc}
  (\bibinfo{year}{2021}).
\newblock \urlprefix\url{https://zenodo.org/record/5256134}.

\bibitem{Finn1992}
\bibinfo{author}{{Finn}, L.~S.}
\newblock \bibinfo{title}{{Detection, measurement, and gravitational
  radiation}}.
\newblock \emph{\bibinfo{journal}{\prd}} \textbf{\bibinfo{volume}{46}},
  \bibinfo{pages}{5236--5249} (\bibinfo{year}{1992}).

\bibitem{Cutler1994}
\bibinfo{author}{{Cutler}, C.} \& \bibinfo{author}{{Flanagan}, {\'E}.~E.}
\newblock \bibinfo{title}{{Gravitational waves from merging compact binaries:
  How accurately can one extract the binary's parameters from the inspiral
  waveform\textbackslash?}}
\newblock \emph{\bibinfo{journal}{\prd}} \textbf{\bibinfo{volume}{49}},
  \bibinfo{pages}{2658--2697} (\bibinfo{year}{1994}).

\bibitem{abbott2020}
\bibinfo{author}{Abbott, B.~P.} \emph{et~al.}
\newblock \bibinfo{title}{Gw190425: Observation of a compact binary coalescence
  with total mass $\sim$ 3.4 m$_{\odot}$}.
\newblock \emph{\bibinfo{journal}{The Astrophysical Journal}}
  \textbf{\bibinfo{volume}{892}}, \bibinfo{pages}{L3} (\bibinfo{year}{2020}).
\newblock \urlprefix\url{http://dx.doi.org/10.3847/2041-8213/ab75f5}.

\bibitem{Gehrels2016}
\bibinfo{author}{{Gehrels}, N.} \emph{et~al.}
\newblock \bibinfo{title}{{Galaxy Strategy for LIGO-Virgo Gravitational Wave
  Counterpart Searches}}.
\newblock \emph{\bibinfo{journal}{\apj}} \textbf{\bibinfo{volume}{820}},
  \bibinfo{pages}{136} (\bibinfo{year}{2016}).

\bibitem{shapiro83}
\bibinfo{author}{{Shapiro}, S.~L.} \& \bibinfo{author}{{Teukolsky}, S.~A.}
\newblock \emph{\bibinfo{title}{{Black holes, white dwarfs, and neutron stars :
  the physics of compact objects}}} (\bibinfo{year}{1983}).

\bibitem{gao20}
\bibinfo{author}{{Gao}, H.} \emph{et~al.}
\newblock \bibinfo{title}{{Relation between gravitational mass and baryonic
  mass for non-rotating and rapidly rotating neutron stars}}.
\newblock \emph{\bibinfo{journal}{Frontiers of Physics}}
  \textbf{\bibinfo{volume}{15}}, \bibinfo{pages}{24603} (\bibinfo{year}{2020}).

\bibitem{rowlinson10}
\bibinfo{author}{{Rowlinson}, A.} \emph{et~al.}
\newblock \bibinfo{title}{{The unusual X-ray emission of the short Swift GRB
  090515: evidence for the formation of a magnetar?}}
\newblock \emph{\bibinfo{journal}{\mnras}} \textbf{\bibinfo{volume}{409}},
  \bibinfo{pages}{531--540} (\bibinfo{year}{2010}).

\bibitem{rowlinson13}
\bibinfo{author}{{Rowlinson}, A.}, \bibinfo{author}{{O'Brien}, P.~T.},
  \bibinfo{author}{{Metzger}, B.~D.}, \bibinfo{author}{{Tanvir}, N.~R.} \&
  \bibinfo{author}{{Levan}, A.~J.}
\newblock \bibinfo{title}{{Signatures of magnetar central engines in short GRB
  light curves}}.
\newblock \emph{\bibinfo{journal}{\mnras}} \textbf{\bibinfo{volume}{430}},
  \bibinfo{pages}{1061--1087} (\bibinfo{year}{2013}).

\bibitem{lv15}
\bibinfo{author}{{L{\"u}}, H.-J.}, \bibinfo{author}{{Zhang}, B.},
  \bibinfo{author}{{Lei}, W.-H.}, \bibinfo{author}{{Li}, Y.} \&
  \bibinfo{author}{{Lasky}, P.~D.}
\newblock \bibinfo{title}{{The Millisecond Magnetar Central Engine in Short
  GRBs}}.
\newblock \emph{\bibinfo{journal}{\apj}} \textbf{\bibinfo{volume}{805}},
  \bibinfo{pages}{89} (\bibinfo{year}{2015}).

\bibitem{Gao2016}
\bibinfo{author}{{Gao}, H.}, \bibinfo{author}{{Zhang}, B.} \&
  \bibinfo{author}{{L{\"u}}, H.-J.}
\newblock \bibinfo{title}{{Constraints on binary neutron star merger product
  from short GRB observations}}.
\newblock \emph{\bibinfo{journal}{\prd}} \textbf{\bibinfo{volume}{93}},
  \bibinfo{pages}{044065} (\bibinfo{year}{2016}).

\bibitem{timmes96}
\bibinfo{author}{{Timmes}, F.~X.}, \bibinfo{author}{{Woosley}, S.~E.} \&
  \bibinfo{author}{{Weaver}, T.~A.}
\newblock \bibinfo{title}{{The Neutron Star and Black Hole Initial Mass
  Function}}.
\newblock \emph{\bibinfo{journal}{\apj}} \textbf{\bibinfo{volume}{457}},
  \bibinfo{pages}{834} (\bibinfo{year}{1996}).

\bibitem{wandermanpiran15}
\bibinfo{author}{{Wanderman}, D.} \& \bibinfo{author}{{Piran}, T.}
\newblock \bibinfo{title}{{The rate, luminosity function and time delay of
  non-Collapsar short GRBs}}.
\newblock \emph{\bibinfo{journal}{\mnras}} \textbf{\bibinfo{volume}{448}},
  \bibinfo{pages}{3026--3037} (\bibinfo{year}{2015}).

\bibitem{metzger17}
\bibinfo{author}{{Metzger}, B.~D.}
\newblock \bibinfo{title}{{Kilonovae}}.
\newblock \emph{\bibinfo{journal}{Living Reviews in Relativity}}
  \textbf{\bibinfo{volume}{20}}, \bibinfo{pages}{3} (\bibinfo{year}{2017}).

\bibitem{shapiro00}
\bibinfo{author}{{Shapiro}, S.~L.}
\newblock \bibinfo{title}{{Differential Rotation in Neutron Stars: Magnetic
  Braking and Viscous Damping}}.
\newblock \emph{\bibinfo{journal}{\apj}} \textbf{\bibinfo{volume}{544}},
  \bibinfo{pages}{397--408} (\bibinfo{year}{2000}).

\bibitem{margalit19}
\bibinfo{author}{{Margalit}, B.} \& \bibinfo{author}{{Metzger}, B.~D.}
\newblock \bibinfo{title}{{The Multi-messenger Matrix: The Future of Neutron
  Star Merger Constraints on the Nuclear Equation of State}}.
\newblock \emph{\bibinfo{journal}{\apjl}} \textbf{\bibinfo{volume}{880}},
  \bibinfo{pages}{L15} (\bibinfo{year}{2019}).

\bibitem{cook94}
\bibinfo{author}{{Cook}, G.~B.}, \bibinfo{author}{{Shapiro}, S.~L.} \&
  \bibinfo{author}{{Teukolsky}, S.~A.}
\newblock \bibinfo{title}{{Rapidly Rotating Neutron Stars in General
  Relativity: Realistic Equations of State}}.
\newblock \emph{\bibinfo{journal}{\apj}} \textbf{\bibinfo{volume}{424}},
  \bibinfo{pages}{823} (\bibinfo{year}{1994}).

\bibitem{breu16}
\bibinfo{author}{{Breu}, C.} \& \bibinfo{author}{{Rezzolla}, L.}
\newblock \bibinfo{title}{{Maximum mass, moment of inertia and compactness of
  relativistic stars}}.
\newblock \emph{\bibinfo{journal}{\mnras}} \textbf{\bibinfo{volume}{459}},
  \bibinfo{pages}{646--656} (\bibinfo{year}{2016}).

\bibitem{lasota96}
\bibinfo{author}{{Lasota}, J.-P.}, \bibinfo{author}{{Haensel}, P.} \&
  \bibinfo{author}{{Abramowicz}, M.~A.}
\newblock \bibinfo{title}{{Fast Rotation of Neutron Stars}}.
\newblock \emph{\bibinfo{journal}{\apj}} \textbf{\bibinfo{volume}{456}},
  \bibinfo{pages}{300} (\bibinfo{year}{1996}).

\bibitem{ai20}
\bibinfo{author}{{Ai}, S.}, \bibinfo{author}{{Gao}, H.} \&
  \bibinfo{author}{{Zhang}, B.}
\newblock \bibinfo{title}{{What Constraints on the Neutron Star Maximum Mass
  Can One Pose from GW170817 Observations?}}
\newblock \emph{\bibinfo{journal}{\apj}} \textbf{\bibinfo{volume}{893}},
  \bibinfo{pages}{146} (\bibinfo{year}{2020}).

\bibitem{li16}
\bibinfo{author}{{Li}, A.} \emph{et~al.}
\newblock \bibinfo{title}{{Internal x-ray plateau in short GRBs: Signature of
  supramassive fast-rotating quark stars?}}
\newblock \emph{\bibinfo{journal}{\prd}} \textbf{\bibinfo{volume}{94}},
  \bibinfo{pages}{083010} (\bibinfo{year}{2016}).

\bibitem{ravi14}
\bibinfo{author}{{Ravi}, V.} \& \bibinfo{author}{{Lasky}, P.~D.}
\newblock \bibinfo{title}{{The birth of black holes: neutron star collapse
  times, gamma-ray bursts and fast radio bursts}}.
\newblock \emph{\bibinfo{journal}{\mnras}} \textbf{\bibinfo{volume}{441}},
  \bibinfo{pages}{2433--2439} (\bibinfo{year}{2014}).

\bibitem{lasky14}
\bibinfo{author}{{Lasky}, P.~D.}, \bibinfo{author}{{Haskell}, B.},
  \bibinfo{author}{{Ravi}, V.}, \bibinfo{author}{{Howell}, E.~J.} \&
  \bibinfo{author}{{Coward}, D.~M.}
\newblock \bibinfo{title}{{Nuclear equation of state from observations of short
  gamma-ray burst remnants}}.
\newblock \emph{\bibinfo{journal}{\prd}} \textbf{\bibinfo{volume}{89}},
  \bibinfo{pages}{047302} (\bibinfo{year}{2014}).

\bibitem{gao16}
\bibinfo{author}{{Gao}, H.}, \bibinfo{author}{{Zhang}, B.} \&
  \bibinfo{author}{{L{\"u}}, H.-J.}
\newblock \bibinfo{title}{{Constraints on binary neutron star merger product
  from short GRB observations}}.
\newblock \emph{\bibinfo{journal}{\prd}} \textbf{\bibinfo{volume}{93}},
  \bibinfo{pages}{044065} (\bibinfo{year}{2016}).

\bibitem{GCN190425z}
\bibinfo{author}{{Ligo Scientific Collaboration}} \& \bibinfo{author}{{VIRGO
  Collaboration}}.
\newblock \bibinfo{title}{{LIGO/Virgo S190425z: Identification of a GW compact
  binary merger candidate.}}
\newblock \emph{\bibinfo{journal}{GRB Coordinates Network}}
  \textbf{\bibinfo{volume}{24168}}, \bibinfo{pages}{1} (\bibinfo{year}{2019}).

\bibitem{INTEGRALGCN3}
\bibinfo{author}{{Minaev}, P.}, \bibinfo{author}{{Pozanenko}, A.},
  \bibinfo{author}{{Grebenev}, S.} \& \bibinfo{author}{{Chelovekov}, I.}
\newblock \bibinfo{title}{{LIGO/Virgo S190425z: INTEGRAL SPI-ACS prompt
  observation.}}
\newblock \emph{\bibinfo{journal}{GRB Coordinates Network}}
  \textbf{\bibinfo{volume}{24170}}, \bibinfo{pages}{1} (\bibinfo{year}{2019}).

\bibitem{Savchenko2012}
\bibinfo{author}{{Savchenko}, V.}, \bibinfo{author}{{Neronov}, A.} \&
  \bibinfo{author}{{Courvoisier}, T.~J.~L.}
\newblock \bibinfo{title}{{Timing properties of gamma-ray bursts detected by
  SPI-ACS detector onboard INTEGRAL}}.
\newblock \emph{\bibinfo{journal}{\aap}} \textbf{\bibinfo{volume}{541}},
  \bibinfo{pages}{A122} (\bibinfo{year}{2012}).

\bibitem{INTEGRALGCN1}
\bibinfo{author}{{Martin-Carrillo}, A.} \emph{et~al.}
\newblock \bibinfo{title}{{LIGO/Virgo S190426c: INTEGRAL prompt observation.}}
\newblock \emph{\bibinfo{journal}{GRB Coordinates Network}}
  \textbf{\bibinfo{volume}{24242}}, \bibinfo{pages}{1} (\bibinfo{year}{2019}).

\bibitem{INTEGRALGCN2}
\bibinfo{author}{{Savchenko}, V.} \emph{et~al.}
\newblock \bibinfo{title}{{LIGO/Virgo S190425z: further analysis of INTEGRAL
  data.}}
\newblock \emph{\bibinfo{journal}{GRB Coordinates Network}}
  \textbf{\bibinfo{volume}{24178}}, \bibinfo{pages}{1} (\bibinfo{year}{2019}).

\bibitem{pozanenko2019}
\bibinfo{author}{Pozanenko, A.~S.}, \bibinfo{author}{Minaev, P.~Y.},
  \bibinfo{author}{Grebenev, S.~A.} \& \bibinfo{author}{Chelovekov, I.~V.}
\newblock \bibinfo{title}{Observation of the second ligo/virgo event connected
  with a binary neutron star merger s190425z in the gamma-ray range}.
\newblock \emph{\bibinfo{journal}{Astronomy Letters}}
  \textbf{\bibinfo{volume}{45}}, \bibinfo{pages}{710–727}
  (\bibinfo{year}{2019}).
\newblock \urlprefix\url{http://dx.doi.org/10.1134/S1063773719110057}.

\bibitem{LVK_CHIME_2022}
\bibinfo{author}{{The LIGO Scientific Collaboration}}, \bibinfo{author}{{the
  Virgo Collaboration}}, \bibinfo{author}{{the KAGRA Collaboration}},
  \bibinfo{author}{{the CHIME/FRB Collaboration}} \emph{et~al.}
\newblock \bibinfo{title}{{Search for Gravitational Waves Associated with Fast
  Radio Bursts Detected by CHIME/FRB During the LIGO--Virgo Observing Run
  O3a}}.
\newblock \emph{\bibinfo{journal}{arXiv e-prints}}
  \bibinfo{pages}{arXiv:2203.12038} (\bibinfo{year}{2022}).

\bibitem{pleunis2021}
\bibinfo{author}{Pleunis, Z.} \emph{et~al.}
\newblock \bibinfo{title}{Lofar detection of 110–188 mhz emission and
  frequency-dependent activity from frb 20180916b}.
\newblock \emph{\bibinfo{journal}{The Astrophysical Journal Letters}}
  \textbf{\bibinfo{volume}{911}} (\bibinfo{year}{2020}).

\bibitem{LVK_merger_rates_2021}
\bibinfo{author}{{The LIGO Scientific Collaboration}}, \bibinfo{author}{{the
  Virgo Collaboration}}, \bibinfo{author}{{the KAGRA Collaboration}}
  \emph{et~al.}
\newblock \bibinfo{title}{{The population of merging compact binaries inferred
  using gravitational waves through GWTC-3}}.
\newblock \emph{\bibinfo{journal}{arXiv e-prints}}
  \bibinfo{pages}{arXiv:2111.03634} (\bibinfo{year}{2021}).

\bibitem{ravi2019}
\bibinfo{author}{Ravi, V.}
\newblock \bibinfo{title}{The prevalence of repeating fast radio bursts}.
\newblock \emph{\bibinfo{journal}{Nature Astronomy}}
  \textbf{\bibinfo{volume}{3}}, \bibinfo{pages}{928–931}
  (\bibinfo{year}{2019}).
\newblock \urlprefix\url{http://dx.doi.org/10.1038/s41550-019-0831-y}.

\bibitem{luo2020}
\bibinfo{author}{Luo, R.} \emph{et~al.}
\newblock \bibinfo{title}{On the frb luminosity function – – ii. event rate
  density}.
\newblock \emph{\bibinfo{journal}{Monthly Notices of the Royal Astronomical
  Society}} \textbf{\bibinfo{volume}{494}}, \bibinfo{pages}{665–679}
  (\bibinfo{year}{2020}).
\newblock \urlprefix\url{http://dx.doi.org/10.1093/mnras/staa704}.

\bibitem{james2022L}
\bibinfo{author}{{James}, C.~W.} \emph{et~al.}
\newblock \bibinfo{title}{{The fast radio burst population evolves, consistent
  with the star formation rate}}.
\newblock \emph{\bibinfo{journal}{\mnras}} \textbf{\bibinfo{volume}{510}},
  \bibinfo{pages}{L18--L23} (\bibinfo{year}{2022}).

\bibitem{Matplotlib2007}
\bibinfo{author}{{Hunter}, J.~D.}
\newblock \bibinfo{title}{{Matplotlib: A 2D Graphics Environment}}.
\newblock \emph{\bibinfo{journal}{Computing in Science and Engineering}}
  \textbf{\bibinfo{volume}{9}}, \bibinfo{pages}{90--95} (\bibinfo{year}{2007}).

\bibitem{Numpy2011}
\bibinfo{author}{{van der Walt}, S.}, \bibinfo{author}{{Colbert}, S.~C.} \&
  \bibinfo{author}{{Varoquaux}, G.}
\newblock \bibinfo{title}{{The NumPy Array: A Structure for Efficient Numerical
  Computation}}.
\newblock \emph{\bibinfo{journal}{Computing in Science and Engineering}}
  \textbf{\bibinfo{volume}{13}}, \bibinfo{pages}{22--30}
  (\bibinfo{year}{2011}).

\bibitem{SciPy2019}
\bibinfo{author}{{Virtanen}, P.} \emph{et~al.}
\newblock \bibinfo{title}{{SciPy 1.0: fundamental algorithms for scientific
  computing in Python}}.
\newblock \emph{\bibinfo{journal}{Nature Methods}}
  \textbf{\bibinfo{volume}{17}}, \bibinfo{pages}{261--272}
  (\bibinfo{year}{2020}).

\bibitem{pandas}
\bibinfo{author}{{W}es {M}c{K}inney}.
\newblock \bibinfo{title}{{D}ata {S}tructures for {S}tatistical {C}omputing in
  {P}ython}.
\newblock In \bibinfo{editor}{{S}t\'efan van~der {W}alt} \&
  \bibinfo{editor}{{J}arrod {M}illman} (eds.)
  \emph{\bibinfo{booktitle}{{P}roceedings of the 9th {P}ython in {S}cience
  {C}onference}}, \bibinfo{pages}{56 -- 61} (\bibinfo{year}{2010}).

\bibitem{reback2020pandas}
\bibinfo{author}{pandas~development team, T.}
\newblock \bibinfo{title}{pandas-dev/pandas: Pandas} (\bibinfo{year}{2020}).
\newblock \urlprefix\url{https://doi.org/10.5281/zenodo.3509134}.

\end{thebibliography}

\newpage

\section{Methods}

\subsection{Search Method}

We search for coincidences between GW signals\cite{abbott2021} and published CHIME FRBs\cite{chime2021}, taking into account all existing theoretical models for potential GW-FRB associations\cite{piro2012,totani2013,zhang2014,wang2016}\citemethods{zhang2020}. Our initial search considers coincidence in time and sky direction, and aims to identify potentially interesting events for further analysis --- distance and dispersion measure are considered only in the combined chance probability calculation\citemethods{Alex2021thesis}.

\paragraph{Time}
Our search time window is chosen to be asymmetrical and 26 hours wide, encompassing FRBs that occur up to 2 hours before a GW signal and 24 hours after. This covers pre-merger emission theories such as magnetospheric interactions\cite{piro2012,wang2016}\citemethods{zhang2016} and magnetic braking\cite{totani2013}, and post-merger emission theories such as magnetar collapse\cite{zhang2014} (note this theory relates only to BNS merger events).  We source our GW data from the published GW catalog GWTC-2 available at the time of writing\cite{abbott2021} 
and CHIME FRBs from the CHIME/FRB Public Database (\url{https://www.chime-frb.ca/catalog}). Our GW sample consists of 39 events detected by LIGO and Virgo during the O3a observing run (April 1, 2019 -- October 1, 2019) and 536 FRBs --- 474 of which are apparent non-repeaters --- published in CHIME Catalog 1\cite{chime2021}. Within our sample, 21 GW events and 171 (150 non-repeating) FRBs have overlapping observing windows and are selected to search for temporal coincidence.
A GW and FRB are considered to be coincident in time if the FRB lies within the designated 26-hr time window of a GW, and this is found by simply iterating over the GPS times of the two signals. Any signals that are not coincident in time are cut, and the remaining candidates are screened for coincidence in sky direction.

\paragraph{Sky Direction} The accuracy of GW localization is mainly determined by the measured signal arrival time between detectors, as well as the relative signal amplitude at each detector\cite{wen2010, fairhurst2011, klimenko2011,  pankow2018}. As a result, it is largely dependent on the number and geographical separation of interferometers that successfully detect a signal. Bayesian methods are used to map the sky direction of a GW source as a posterior probability distribution on the sky\citemethods{singer2016}, with a `credible interval' value being assigned to each RA and DEC coordinate. We obtain these credible interval values from the published GWTC-2 data\cite{abbott2021}. 
 Note a smaller credible level value indicates
 closer proximity to the optimal 
 value\citemethods{sivia2006}.  The current sensitivities of the Advanced LIGO and Virgo network are able to localize GW signals to sky areas of tens of square degrees\cite{abbott2020b} for confident detections by all three interferometers. CHIME FRBs are localized using beam model predictions with errors on the order of arcminutes\cite{chime2021}. Therefore, for our sky direction cut, we consider a GW-FRB pair spatially coincident if the best-fitting FRB coordinates lie within the 90\text{\%} percent credible interval of candidate GW signals. This is done using the 
\texttt{find\_greedy\_credible\_levels} function of the \texttt{ligo.skymap.postprocess.util} module\citemethods{ligoskymap}.

\subsection{Search Results}
Our search found $12$ out of $21$ ($\sim 57\%$) GW events coincident in time with at least one CHIME FRB. This high fraction is not surprising considering the high rate of CHIME FRB detections (see ``Chance Probability Derivation'' section). One GW event initially included in GWTC-2 that was temporally coincident with two CHIME FRBs --- GW190424\_180648 --- has since been redacted due to its significance being reduced upon further re-analysis and is not included in any updated GWTC event lists (https://www.gw-openscience.org/GWTC-2.1).
Thus, we eliminate this candidate coincidence from the spatial search. Of the remaining GW-FRB pairs, only one candidate coincidence passed the sky direction cut: FRB 20190425A - GW190425. Remarkably, this is also the only BNS event detected during O3a.

\subsection{Chance Probability Derivation}

Given the wide time window we are searching over, the high detection rate of CHIME FRBs and the poor localization of GW events, random coincidences in time and sky direction are expected. Furthermore, since the relation between dispersion measure (DM) and redshift $z$ for FRBs shows large fluctuations\cite{macquart2020} --- and due to the uncertainty in $z$ inferred from GW190425 itself --- we have not placed any a priori cuts on the predicted DM values from the estimated luminosity distance of GW190425. Nonetheless, FRBs arriving closer in time, from a sky direction consistent with the GW event localization, and with DMs consistent with the inferred GW event distance given the ionized matter content of the Universe, will present greater evidence against such a chance association. We derive a joint probability of chance association by combining probabilities of random temporal, spatial and DM (or distance) coincidences defined as $P_{\rm T}$, $P_{\rm S}$, and $P_{\rm DM}$, respectively. As non-repeating FRBs best suit compact object merger theories, we choose to analyse the distributions of non-repeating CHIME FRBs only, though a conservative value is also calculated using all CHIME FRBs (i.e.\ including repeaters).
Details of the calculation are below.

\paragraph{$P_{\rm T}$}

\begin{figure}
    \centering
    \includegraphics[width=\textwidth]{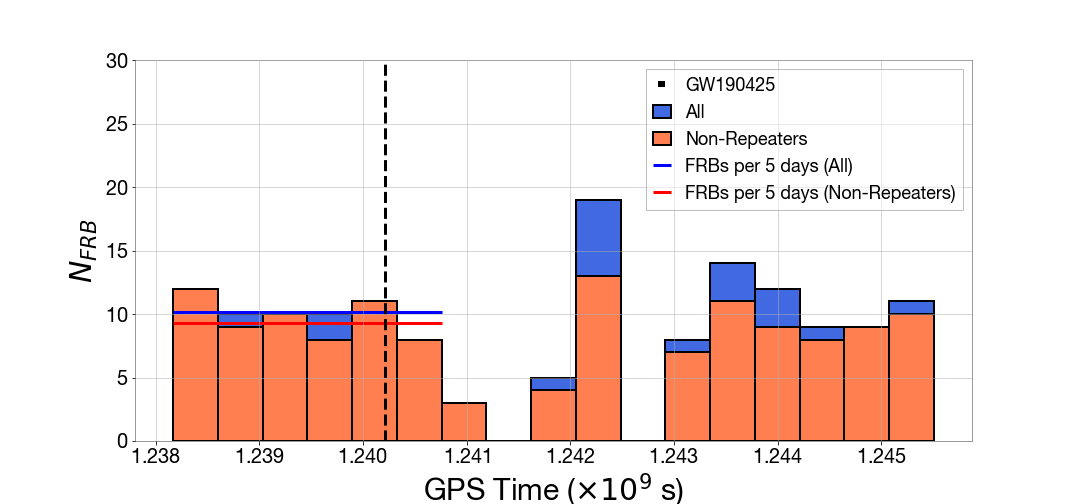}
    \caption{{\bf CHIME FRB detection rates.} Number $N_{\rm FRB}$ of all (blue histogram) and non-repeating (orange histogram) CHIME FRBs per 5 days. Average rate during the period surrounding GW190425 (dashed black line) for all and non-repeating FRBs shown by the blue and red lines, respectively.}
    \label{fig:frb_detection_times}
\end{figure}

The likelihood of a chance temporal coincidence, $P_{\rm T}$, is the probability of CHIME randomly detecting an FRB in the 2.5\,hour post-merger, as per the blitzar scenario. GW190425 was the only BNS event detected during the overlap of O3a and CHIME Catalog 1 --- see P-Value Calculation for calculations considering other scenarios, e.g.\ BH--BH mergers.
CHIME's operationality was not constant throughout O3a, during which CHIME's FRB detection efficiency was affected by scheduled maintenance and unscheduled outages\cite{chime2021}. In particular, the detection rate reduces for a short period in the days after GW190425. As a result, we cannot assume a uniform distribution of CHIME FRB detections. To derive a more accurate value of the rate of FRBs ($R_{\rm FRB}$) around the time of GW190425, we group CHIME FRBs into 5-day bins, and use the time-averaged number of detections within a 30-day period (see \EXTFIG{fig:frb_detection_times}) to find $R_{\rm FRB}$. Thus we average over variability in CHIME sensitivity on small timescales. The resulting rates are 1.93 FRBs per day for non-repeaters, and 2.03 per day when including repeaters. The number of FRBs expected to fall within a 2.5 hour time window is then 
$2.5 \cdot 1.93/24 = 0.20$  for non-repeaters and $2.5 \cdot 2.03/24 = 0.21$ for our conservative value considering all CHIME FRBs.

The number of once-off FRB detections in any given time period $T$ will follow a Poissonian distribution with expectation value $\lambda=T R_{\rm FRB} \approx 0.20$ for a 2.5 hour window. The probability $P_{\rm T}$ to observe a random CHIME non-repeating FRB with a tighter time coincidence than FRB~20190425A is then given by:
\begin{eqnarray}
P_{\rm T} = 1-\exp(-\lambda) \approx 0.18. \label{eq:ptime}
\end{eqnarray}
for non-repeating FRBs. The periodic and/or bursty activity periods for repeating FRBs will make the time distribution from these sources non-Poissonian\citemethods{CHIME2020_periodic,Rajwade_121102_repetitions}. Ignoring the non-Poissonian nature of repeating FRBs, and using $\lambda \approx 0.21$ for the coincidence window, produces a conservative $P_{\rm T} = 0.19$.

If instead we consider only our search time window of $T=26$\,hr, the expected rate $\lambda=2.1$, and hence $P_{\rm T}=0.88$.

\paragraph{$P_{\rm S}$} We define $P_{\rm S}$ to be the probability that a random CHIME FRB would be detected at a location with equal to or higher likelihood in the GW190425 skymap than FRB 20190425A (i.e.\ within the 66.7\% credible interval). 
This is a function of GW190425's refined skymap\citemethods{abbott2021}, weighted by the exposure time of CHIME at each coordinate. This can be written as:
\begin{eqnarray}
    \label{eq:spatial}
    P_{\rm S} & = & \int_{\Omega_x} E(\Omega)d\Omega,
\end{eqnarray}
where $\Omega$ is a general 2D sky coordinate, $\Omega_x$ is the 66.7\text{\%} credible region and $E(\Omega)$ is the relative likelihood (`exposure') of seeing an FRB at sky position $\Omega$.

CHIME's instantaneous exposure is a function of its primary beamshape and CHIME's location at a latitude of $49^\circ 19'13''.08$ North\cite{chime2021}. We ignore declination dependent fluctuations due to the placement of the synthesized beams, which are on scales much smaller (${\mathcal O}\sim0.1^\circ$) than variations in the GW190425 skymap. We also approximate the instantaneous coverage in local azimuth angle to be very small, which is valid at the declinations considered here. Therefore, any background FRB detected in the [0,2.5] hour window after GW190425 --- i.e.\ the 2.5 hour window before FRB 20190425A --- will necessarily fall in the RA ($\alpha$) range $218.22^{\circ} \le \alpha \le 255.72^{\circ}$. This method implicitly accounts for the favourable alignment of the LIGO and CHIME antenna patterns due to their nearby geographical location, which causes the GW190425 skymap to significantly overlap with the sky area covered by CHIME's primary beam within the 2.5 hr time window.

The CHIME/FRB Collaboration use a sophisticated pulse-injection method --- accounting for beamshape effects --- to calculate the exposure $E(\delta)$ to an FRB as a function of declination $\delta$ \cite{chime2020}. This exposure time
is equal to the detection probability per square degree to within a normalisation factor, and is split into `upper' and `lower' transit curves representing sky regions viewed above and below the Northern pole. The un-normalized exposure, $E_{u/l}'(\delta)$, is modelled by manually fitting a spline to the upper `u' and lower `l' transit curves (Figure 5 in Ref.\ \cite{chime2021}) of this function. The normalization constant, $C$, is then derived by integrating $E'(\Omega)$ over
all declinations and the RA range of interest using the upper transit curves, and the RA range offset by 12\,hr for the lower transit curves, i.e.\ 
\color{black}
\begin{eqnarray}
    \label{eq:spatial2}
    C = \int_{218.22^{\circ} \le \alpha \le 255.72^{\circ}} E_u'(\delta) d \Omega 
    + \int_{38.22^{\circ} \le \alpha \le 75.72^{\circ}} E_l'(\delta) d \Omega \approx 1658 
\end{eqnarray}
\color{black}

Using our derived value for C, we then integrate the normalized exposure function, $E_{u/l}(\Omega) = E_{u/l}'(\delta)/C$,  over the 66.7\text{\%} credible sky area in the range $218.22^\circ \le \alpha \le 255.72^\circ$ ($38.22^\circ \le \alpha \le 75.72^\circ$) for the upper (lower) transit respectively as per Eq.~\eqref{eq:spatial} to find the total probability. This produces $P_{\rm S} = 0.265$. As a simple check of this method, we also took the optimal measured sky positions of all non-repeating CHIME FRBs, and find that of the 56 in this RA range, 17 of them (30\%) lie within the credible interval. If we instead randomise the RA of all non-repeating CHIME FRBs uniformly over the RA range,
we find that 29.4\% lie within the 66.7\text{\%} credible sky area of GW190425's localization; while shuffling FRB RA values in the CHIME catalog with respect to DEC finds 28.8\% lie in this interval. We attribute the excess beyond our estimates to be random fluctuations of the observed RA,DEC of FRBs compared to expectations.

If we instead consider an FRB uniformly distributed in time over our [-2,+24]\,hr time window (for which $P_{\rm T}=0.88)$, we find $P_S=0.154$ for the 90\% likelihood region. Thus the total chance of an event passing our initial selection criteria is $P_{\rm pass}=$13.5\%.

We observe that the distribution of CHIME FRBs has been extensively analysed for deviations from a simple declination dependent sensitivity\cite{CHIME2021_lat_dependence}, with the only observed anisotropy not due to $E(\delta)$ being a potential correlation between FRBs with DM $\sim$ 800\,pc\,cm$^{-3}$ and large-scale structure at $z\sim0.4$ \cite{CHIME2021_structure}. The spatial scales of such structure are small compared to the variation in the GW190425 skymap, and the distances involved are much larger than those relevant to GW190425, so we ignore this here.

\subsection{P-value calculation}
\label{sec:pvalue}

We use the product $P_{\rm T} P_{\rm S} P_{\rm DM}$ as evidence against our null hypothesis, $H_0$, of a purely chance association. To estimate the probability of obtaining an equally small product under $H_0$ (i.e.\ the p-value),
we consider the 88\% probability of observing an FRB in the time range [-2,24]\,hr about the time of GW190425.  We use a uniform time distribution, with RAs located at the centre of the CHIME beam, and DECs distributed according to CHIME's exposure. We calculate $P_{\rm S} P_{\rm DM}$ for all such FRBs, rejecting those outside the 90\% credible area. Sampling $P_{\rm DM}$ uniformly from $0$ to $1$ allows us to calculate the probability of observing a value of $P_{\rm T} P_{\rm S} P_{\rm DM} < 1.9 \cdot 10^{-4}$ under $H_0$. We find a p-value of 0.0052, i.e.\ the significance of this event is $2.8\,\sigma$.

BNS are the most commonly proposed merger scenario for FRB progenitors \cite{platts2019,LVKchime2022}. However, pre-merger emission from the inspiral of binaries involving a charged black hole has also been proposed\citemethods{Zhang2016bbh}. In such a scenario, the possibility of a CHIME FRB occurring during the [-2,0]\,hr time window prior to the other 20 GW events should be considered as a trial factor. For this calculation, we use the mean CHIME FRB detection rate of 1.62\,FRBs/day during the overlap period between O3 and the CHIME catalogue (138 once-off FRBs in 85 days), for an expected number of 2.7 coincident FRBs. Using the GWTC 2.1 skymaps, and the same methods to calculate $P_S$ above, we find that 0.41 of these would also be expected to pass our spatial selection criteria. Compared to the 13.5\% chance of a background event passing these criteria for GW190425, this effectively counts as three extra trials. Thus if this scenario is considered equally plausible --- which it is not, due to the incredibly large required charge ($3.3 \times 10^{21}$\,C M/M$_{\odot}$) on the BHs --- our p-value should be raised to 0.021, i.e.\ 2.3$\sigma$.

\paragraph{Notes on GW190425 Localization} 
The sky localization of GW190425 is poorly constrained due to it being detected significantly in only a single detector. FRB 20190425A lies within the 66.7\% credible interval for the most recently published GWTC-2 skymap for GW190425\cite{abbott2021} (used in our main analysis). The FRB sky direction is also largely consistent with all skymaps available for the GW event at various stages of its discovery (\EXTTAB{tab:credible}). It has a better consistency with the initial online rapid Bayestar localization\citemethods{singer2016}; using this skymap results in a 50\% lower sky coincidence probability ($P_S$) for a random spatial coincidence than the GWTC-2 result. Our FRB, however, is less consistent with the online LALInference skymap\citemethods{lalinference}. Using this LALInference skymap results in a larger $P_S$, by 32\%. 
The GWTC-2 skymap is obtained using cleaned data and a better estimation of the noise level, while the initial Bayestar and LALInference skymaps are obtained using the available online data around the time of the detection. Therefore we conclude that $P_S=0.265$.

\begin{table}[h]
    \caption{Credible interval that FRB 20190425A lies in for each of the published skymaps for GW190425, and corresponding probabilities for coincidence in sky direction ($P_S$, see text). All skymaps were sourced from
    \url{https://gracedb.ligo.org/superevents/S190425z/view/ }.
    }
    \centering
    \label{tab:credible}
    \begin{tabular}{c|c|c|c}
    \hline\hline
      \textbf{Localization
      } & \textbf{Credible Interval} & \textbf{$P_{S}$} & \textbf{Date Published}\\
      \hline
      Bayestar & 25.0\% & 0.14 & Apr 2019 \\
      \hline
      LALInference & 90.2\% & 0.35 & Apr 2019\\
      \hline
      Updated Superevent & 67.6\% & 0.21 & Jul 2020\\
      \hline
      GWTC-2 & 66.7\% & 0.265 & Oct 2020\\
      \hline
    \end{tabular}
\end{table}

\paragraph{$P_{\rm DM}$} An FRB produced at a given redshift, $z$, will have a likelihood distribution $P({\rm DM|z})$ of dispersion measures, DM. The DM is a measure of the column density of free electrons, $n_e$, along the line of sight $\ell$ from the FRB source to the observer, in units of pc\,cm$^{-3}$. Accounting for the redshift of an extragalactic object, it is defined as
\begin{equation}
    \label{eq:DM}
    {\rm DM} = \int \frac{n_e(\ell)}{1+z}\ d\ell.
\end{equation}
The DM of extragalactic FRBs can be characterised in multiple ways. Here, we use four components: two from the Milky Way, being its interstellar medium (MWISM) and halo (MWhalo), and two extragalactic contributions, from the intergalactic medium (IGM), and the host galaxy (host), two of which depend on redshift $z$:
\begin{equation}
    \label{eq:DMbudget}
    {\rm DM}(z) = {\rm DM}_{\rm MW} + {\rm DM}_{\rm EG}(z) = {\rm DM}_{\rm MWISM} + {\rm DM}_{\rm MWhalo} + {\rm DM}_{\rm IGM}(z) + \frac{{\rm DM}_{\rm host}}{1+z}.
\end{equation}
We observe that in the CHIME Catalog 1 data, quoted `extragalactic' values for ${\rm DM}_{\rm EG}$ include ${\rm DM}_{\rm MWhalo}$, i.e.\ ${\rm DM}_{\rm EG} = {\rm DM}-{\rm DM}_{\rm MWISM}$. The contribution from the ISM, ${\rm DM}_{\rm MWISM}$, can be estimated using various electron density models, e.g.\ NE2001\citemethods{cordes2002}, YMW16\citemethods{yao2017}, or YT20\citemethods{yamasaki2020}, giving 48.79, 38.8 and 49.0 pc\,cm$^{-3}$ respectively for FRB 20190425A. Such models can suffer from over-fitting however, and may have poor predictive power\citemethods{Schnitzeler2012DM}. ${\rm DM}_{\rm MWhalo}$ is uncertain --- lower limits can be placed using pulsars from the Large Magellanic Cloud\citemethods{Shannon2018}, while upper limits can be placed using nearby FRBs\citemethods{CHIME_M81_2021}. Modelling\citemethods{ProchaskaZheng2019} suggests ranges from 50--80 pc\,cm$^{-3}$.

The extragalactic contributions, ${\rm DM}_{\rm IGM} + {\rm DM}_{\rm host}$, have been estimated by fits to five localized FRBs\cite{macquart2020}. Both contributions are expected to have tails towards high DM values,
from the few that intersect many or dense clumps of matter along the line-of-sight, and/or originate from dense regions within their host. Therefore DM values only poorly constrain the distance to an FRB.
However, a robust upper limit, $z_{\rm max}$, can be estimated by setting ${\rm DM}_{\rm host}=0$, taking the lowest of model values for $\rm DM_{\rm MWISM}$ (38.8\,pc\,cm$^{-3}$) and ${\rm DM}_{\rm MWhalo}$ (50\,pc\,cm$^{-3}$), and using the approximate relation\citemethods{Inoue2004} in the nearby Universe of $\rm DM_{\rm IGM}=1000\, {\rm pc\,cm}^{-3}\,z$. Such an estimate is consistent with the nearby low-DM FRB discovered in M81\cite{CHIME_M81_2021}. Applying this to FRB 20190425A produces $z_{\rm max}=0.0394$, or \ $\sim$180\,MPc using standard cosmology\citemethods{ade2016,Wright_2006_cosmology}.

Deriving a realistic DM--z relation using the model described in \cite{macquart2020} requires considering variations in 
the component DM contributions.
We use the NE2001 model for ${\rm DM}_{\rm MWISM}$, and assume ${\rm DM}_{\rm MWhalo}=50$\,pc\,cm$^{-3}$. In
this model, errors in ${\rm DM}_{\rm MWISM}$ and ${\rm DM}_{\rm MWhalo}$ are absorbed into the distributions for ${\rm DM}_{\rm IGM}$ and ${\rm DM}_{\rm host}$.
In their model, ${\rm DM}_{\rm IGM}$ is a function of the total baryon content of the Universe, $\Omega_b$, and a `feedback' parameter F governing how clumped those baryons are\cite{macquart2020}. This DM model is also explained in \citemethods{James2021}, and implemented in publicly available Python code\citemethods{frb,zdm}.

The feedback parameter F is poorly constrained by localized FRBs\cite{macquart2020} --- here, we somewhat arbitrarily use F $=0.32$ --- while the baryon content was consistent with expectations from measurements of the cosmic microwave background\citemethods{PlanckCosmology2018}. Most importantly for this work, fits to the FRB host contribution ${\rm DM}_{\rm host}$ (which dominates over  ${\rm DM}_{\rm IGM}$ in the nearby Universe) used a log-normal distribution,
\begin{eqnarray}
p({\rm DM}_{\rm host}|\mu_{\rm host},\sigma_{\rm host} ) = \frac{1}{{\rm DM_{\rm host}}} \frac{1}{\sigma_{\rm host} \sqrt{2 \pi}} 
\exp \left[ -\frac{(\log {\rm DM}_{\rm host}-\mu_{\rm host})^2}{2 \sigma_{\rm host}^2} \right] \label{eq:phost}.
\end{eqnarray}
Best-fit values of $e^{\mu_{\rm host}}=65-70$ \,pc\,cm$^{-3}$ and $\sigma_{\rm host}=0.4$ (where $\mu_{\rm host}$ and $\sigma_{\rm host}$ are the mean and standard deviation of the fitted lognormal FRB host galaxy DM distributions, respectively) are obtained using a ``gold standard'' sample of five localized FRBs measured with the Australian Square Kilometre Array Pathfinder (ASKAP).

\begin{figure}
    \centering
    \includegraphics[width=0.32\textwidth]{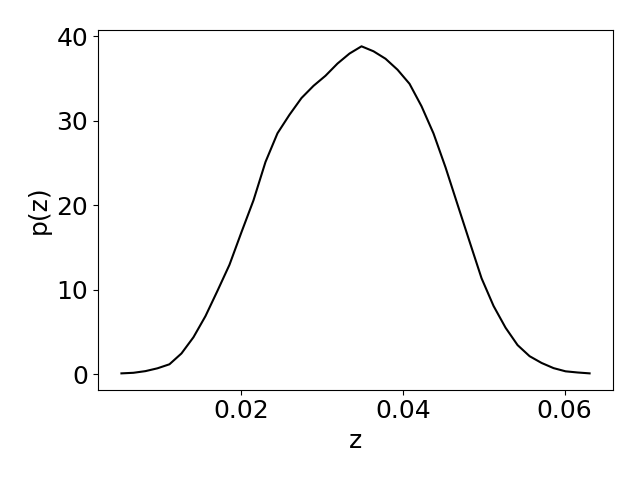}
    \includegraphics[width=0.32\textwidth]{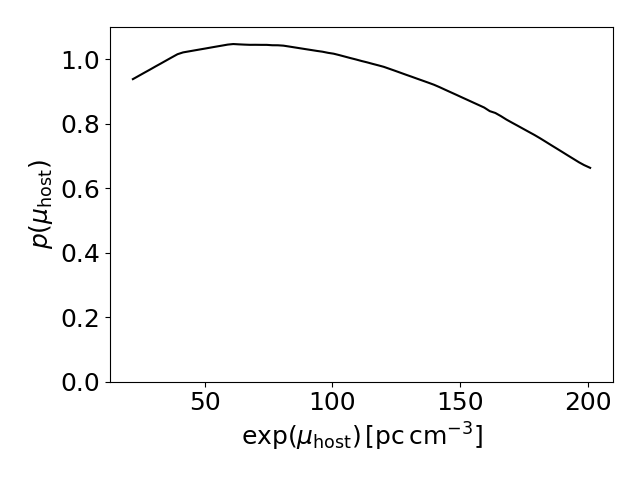}
    \includegraphics[width=0.32\textwidth]{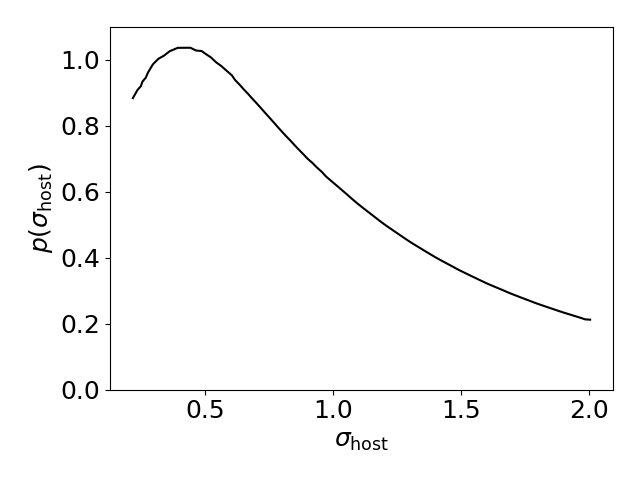}
    \caption{From left to right: probability distribution of redshift $z$ for GW190425 (from \url{https://dcc.ligo.org/LIGO-P2000223/public}); and the probability distributions of the mean and standard deviation of the fitted lognormal FRB host galaxy DM distribution\cite{macquart2020}.}
    \label{fig:uncertainties}
\end{figure}

Incorporating uncertainties in $z$ from GW190425, $\mu_{\rm host}$, and $\sigma_{\rm host}$ (shown in \EXTFIG{fig:uncertainties}), the expected DM of an FRB originating from GW190425 is therefore:
\begin{equation}
    p({\rm DM|GW190425}) = \int p({\rm DM}|z,\mu_{\rm host},\sigma_{\rm host}) p(z) p(\mu_{\rm host}) p(\sigma_{\rm host}) dz d\mu_{\rm host} d\sigma_{\rm host}, \label{eq:pdm}
\end{equation}
where $p({\rm DM}|z,\mu_{\rm host},\sigma_{\rm host})$ is derived from \eqref{eq:DMbudget} and \eqref{eq:phost}.
Integrating over these distributions produces the expected DM distribution. We plot in Figure~\ref{fig:expected_DM} the expected distribution with the DM$_{\rm MWISM}$ contribution subtracted, to allow comparison with other once-off CHIME FRBs\cite{chime2021}.

We calculate the coincidence probability associated with the DM of FRB 20190425A by ordering all 474 non-repeating CHIME FRBs according to $p($DM-DM$_{\rm MWISM})$ and counting the number with equal or better probabilities. Using the NE2001 model for DM$_{\rm MWISM}$, we find only one FRB with a better-matching $p($DM-DM$_{\rm MWISM})$. Thus, a simple estimation yields $P_{\rm DM}=2/474 \approx 0.0042$. 
The uncertainties in DM$_{\rm MWISM}$,  $\mu_{\rm host}$ and $\sigma_{\rm host}$ have also been derived using NE2001\cite{macquart2020}, i.e.\ using NE2001 only is self-consistent. Nonetheless, we calculate a conservative value of $P_{\rm DM}$ using two methods. Firstly, we simply use the YMW16 model to calculate DM$_{\rm MWISM}$ for all CHIME FRBs. In this case we find four other FRBs with a more probable DM value than FRB 20190425A, i.e.\ $P_{\rm DM}=5/474 \approx 0.0105$. We also simulate uncertainty in DM$_{\rm MWISM}$ by randomly generating these values from a Normal distribution with mean equal to the average of NE2001 and YMW16, and standard deviation equal to the difference. Out of 1000 random iterations, we find on-average 4.2 more-probable FRBs, i.e.\ $P_{\rm DM}=5.2/474 \approx 0.011$. We take this value as a conservative estimate of $P_{\rm DM}$.

Therefore $P_{\rm tot} = P_T P_S P_{\rm DM} = 0.18 \cdot 0.265 \cdot 0.004\ (0.19 \cdot 0.265 \cdot 0.011) = 1.9 \cdot 10^{-4}\ (5.5 \cdot 10^{-4})$ is the total (conservative) chance probability. These probabilities were derived using a Frequentist approach, but we note that a Bayesian measure of significance can also be applied\citemethods{ashton2018}.

\subsection{GW190425 SNR time series} We use modules from the \texttt{PyCBC} software package\citemethods{Nitz2021} to perform a matched filtering\citemethods{Finn1992,Cutler1994} and obtain the SNR time series in Figure \ref{fig:skymap}. Public glitch subtracted, cleaned  GW strain data is used for the filtering (\url{https://dcc.ligo.org/LIGO-T1900685/public}) 
together with a \texttt{TaylorF2} waveform template for the most probable chirp mass of $1.48M_\odot$
\cite{abbott2020}.

\subsection{Host galaxy association search}

\begin{table}
	\begin{tabular}{ccccccc}
		\hline
		Name & z & RA & DEC \\
		\hline
	    WISEA J170311.62+212626.6 & 0.078742 & 17 03 11.622 & +21 26 26.61  \\
	    UGC 10667 & 0.031 & 17 02 38.976 & +21 34 35.91\\
	    WISEA J170310.07+212309.9 & 0.047523 & 17 03 10.07 & +21 23 09.9 \\
	    
		\hline
	\end{tabular}
	\caption{List of extra-galactic sources within the central 68\% localization uncertainty of FRB20190425 ($0.1\times0.2\deg^2$) retrieved via the NASA Extragalactic Database (NED) within a redshift range $0.001 < z < 0.08$ consistent with the GW PE redshift distribution. Quoted redshifts are spectroscopic. We exclude objects that do not have a listed spectroscopic redshift.
	}
	\label{tab:galsincircle}
\end{table}

The properties of potential host galaxies found inside the error ellipse of FRB 20190425A's central localization are displayed in \EXTTAB{tab:galsincircle}. The only host within the upper limit $z_{max} = 0.0394$ is UGC 10667. To investigate the significance of this host galaxy association we utilize the methods described in \cite{Gehrels2016}. The total number of galaxies expected within a solid angle $\Delta \Omega$ at luminosity distance $d_\mathrm{L}$ is $N_\mathrm{gals}=\rho_\mathrm{gals} \frac{4}{3}\pi d_\mathrm{L}^3 \frac{\Delta \Omega}{4 \pi}$ in the nearby (Cartesian) Universe. The density of galaxies can be derived from the Schechter function associated with a galaxy catalog optimized for GW follow-up searches.\cite{Gehrels2016} In this work we utilize their value of $\rho_\mathrm{Gals} = 2.35 \cdot 10^{-3}\,\mathrm{Mpc}$, which considers galaxies that contribute to the top 50\% of the luminosity function. Thus, for a  luminosity distance of $255.85\,\mathrm{Mpc}$ (the upper limit derived for the GW event\cite{abbott2020}) one expects $N_\mathrm{Gals}\sim  0.173$ within the CHIME 68\% FRB 20190425 error ellipse (i.e. $\sim 17\%$ chance of coincidence).

\subsection{Delay time}
Within the ``blitzar'' model for FRBs, the delay time between GW190425 and FRB 20190425A is defined by the survival time of the supramassive neutron star (SMNS) formed at the merger before it collapsed into the black hole. The collapse time scale depends on the mass of the SMNS. 
For the majority of the cases, the collapse time likely coincides with the spindown timescale, which, for a magnetic-dipole-dominated spindown\citemethods{shapiro83}, can be estimated as
\begin{eqnarray}
t_{\rm sd,md} = \frac{E_{\rm rot}}{L_{\rm sd,md}} = 3.0 \times 10^4~s~ \left(\frac{R_{NS}}{18 ~{\rm km}}\right)^{-6} \left(\frac{B_p}{10^{14}~{\rm Gauss}}\right)^{-2} \left(\frac{P_i}{0.8 ~{\rm ms}}\right)^2\left(\frac{I}{8\times 10^{45}~{\rm g~cm^2}}\right),
\label{eq:t_spindown}
\end{eqnarray}
where $E_{\rm rot} = (1/2) I \Omega_i^2$ is the total rotational kinetic energy of the SMNS, $L_{\rm sd,md} = B_p^2 R_s^6 \Omega_i^4 / (6c^3) $ is the spindown luminosity due to magnetic dipole radiation, $\Omega_i$ and $P_i$ represent the initial angular velocity and period of the new-born SMNS, $B_p$ stands for the strength of the surface magnetic field at the pole, and $I$ is the momentum of inertia of the SMNS. These parameters have been normalized to typical values at the maximum spin for an SMNS\cite{gao20}. 
For these nominal parameters, a collapse time at $2.5$ hr corresponds to $B_p \sim 1.8 \times 10^{14}$ G.

The light curves of a fraction of short GRBs are found to exhibit an X-ray plateau followed by a steep decay \citemethods{rowlinson10,rowlinson13,lv15}. These ``internal plateaus'' are 
proposed to form from the wind emissions of an SMNS that end with the collapse of the SMNS\cite{zhang2014}. The distribution of the observed plateau duration
falls within the range of $10^2-10^4$s\citemethods{lv15,Gao2016},
consistent with our 
interpretation of the FRB delay time.

\subsection{Constraints on $M_{\rm TOV}$}

We derive constraints on the universal Tolman-Oppenheimer-Volkoff non-spinning maximum mass ($M_{\rm TOV}$) by constraining the gravitational mass of the merger remnant at different post-merger phases. To achieve this, we first derive the total post-merger baryonic mass (a conserved quantity). According to GW parameter estimation\cite{abbott2021}, the gravitational masses of the binary neutron stars in the source frame are $M_1 = 2.03^{+0.58}_{-0.34} M_{\odot}$ and $M_2 = 1.35^{+0.26}_{-0.26}M_{\odot}$ with the sum of the two $M_{\rm tot} = 3.39^{+0.31}_{-0.11}M_{\odot}$. The remnant mass extracted from the GW observation is $M_{\rm tot}^{\rm final} = 3.23^{+0.33}_{-0.11}M_{\odot}$ in the source frame (\url{https://dcc.ligo.org/LIGO-P2000026/public}). This is smaller than $M_{\rm tot}$, indicating a fraction of its mass was radiated away in GWs. For pre-merger NS, we use the non-spin or low-spin universal relation to estimate the total baryonic mass $M_b$ of the system from its gravitational mass $M$: $M_b = M + 0.080M^2$, which is applicable
to NSs that are not spinning near the break-up limit
(ref.\ \citemethods{timmes96,gao20}). We can safely assume low-spins for pre-merger NSs, as 
the spindown timescale (Equation \ref{eq:t_spindown}) of a millisecond object is about $1.3 \times 10^9~{\rm yr}~(B_p/10^8~{\rm Gauss})^{-2}$. Therefore, even if the initial spin of a BNS merger member is close to milliseconds, it should have been spun down during the typical BNS merger timescale $\sim 10^9$ yr\citemethods{wandermanpiran15},
unless its $B$ field is lower than a few $10^8$ G, which is very rare and never observed in the Milky Way.
This gives the total baryonic mass $M_{\rm b,tot} = 3.86_{-0.31}^{+0.70}M_{\odot}$, including $0.1M_{\odot}$ uncertainty introduced by the $M_b - M$ relation due to the unknown neutron star EoS. During the merger, an order of $0.06M_{\odot}$ baryonic mass is expected to have been ejected to power kilonova emission\cite{metzger17}, therefore the baryonic mass of the final remnant may be estimated as $M_{\rm b,rem} = 3.80_{-0.31}^{+0.70}M_{\odot}$.

A neutron star post-merger remnant is expected to go through a brief differential-rotation phase (< 1 second), before forming an essentially uniformly rotating body\cite{shapiro00,margalit19}.
We therefore apply a universal relation between baryonic mass $M_b$ and gravitational mass $M$ for maximally rotating neutron stars, $M_b = M + 0.064 M^2$, to convert the baryonic remnant mass $M_{\rm b,rem}$ to gravitational remnant mass $M_{\rm rem}$ (ref.~\citemethods{gao20}).
This leads to $M_{\rm rem} = 3.16^{+0.40}_{-0.24}M_{\odot}$. Note that $M_{\rm rem}$ is slightly lower than $M^{\rm final}_{\rm tot}$ as $M^{\rm final}_{\rm tot}$ is not necessarily taken from the rigidly rotating phase.
Since uniform rotation can support a higher mass with a maximum  enhancement factor
$M_{\rm rem}/M_{\rm TOV} \sim 1.2$ \cite{cook94,breu16}\citemethods{lasota96} (more precisely, $1.201 \pm 0.017$ \cite{ai20}), a lower limit of $M_{\rm TOV} > 2.63^{+0.39}_{-0.23}M_{\odot}$ can be placed. 
The minimum enhancement factor at the maximum rotation has been shown to be $1.046 \pm 0.008$ \cite{ai20}. This places a limit of $M_{\rm TOV} < 3.02^{+0.42}_{-0.25}M_{\odot}$.  

If a quark star is formed after the merger, due to the lack of a universal $M_b - M$ relation, the gravitational mass of the remnant is difficult to estimate. In this case, we use the final mass $M_{\rm tot}^{\rm final}$ extracted from the GW waveform as the remnant mass $M_{\rm rem}$ to constrain $M_{\rm TOV}$. For rotating quark stars, the enhancement factor can be as high as $M_{\rm rem}/M_{\rm TOV} \sim 1.4$\cite{li16}. On the other hand, since there is no detailed calculation of the enhancement factor for maximum rotation, we adopt the enhancement factor $\sim 1$ to represent a most conservative constraint on the upper limit of $M_{\rm TOV}$ of quick stars. We then find that
$2.31_{-0.08}^{+0.24}M_{\odot}<M_{\rm TOV}<3.23^{+0.33}_{-0.11}M_{\odot}$. 

In principle, $M_{\rm TOV}$ can be further constrained given the 2.5 hour collapse time if the surface magnetic field strength of the post-merger compact star is known\cite{ravi14}. 
The combination of short GRB X-ray plateau observations and theory suggests a collapse time ranging from minutes to hours\cite{lasky14,gao16,li16}. The observed 2.5 hours is consistent with this range.
For our case, if the magnetic field is very low, the increase of spin period would be insignificant in 2.5 hours, in which case a $M_{\rm TOV}$ close to the lower limit is required to ensure the remnant collapse after the star is slightly spun down; If the surface magnetic field is very high, most of the angular momentum would have been lost in a short time. Then a $M_{\rm TOV}$ close to the upper limit is required to prevent the supramassive compact object from collapsing before 2.5 hours. In reality, the surface magnetic field of the merger remnant is not well constrained. Therefore, the allowed $M_{\rm TOV}$ range remains broad, as seen in the dark grey region of Figure \ref{fig:mtov}.

\subsection{Energetics}

The estimated fluence of FRB~20190425A is $31.6 \pm 4.2$\,Jy\,ms\cite{chime2021}. This is a lower limit, and is estimated assuming it occurs at CHIME's beam centre. Using the CHIME bandwidth of $400$\,MHz, and assuming a luminosity distance of $156\,\mathrm{Mpc}$, we calculate an isotropic equivalent energy of $E_\mathrm{FRB}=3.72 \cdot 10^{38}\,\mathrm{erg}$. The emitted FRB power is a small/negligible fraction of the mass-energy of the remnant in the optimistic scenario. In contrast, the GW mass-energy emitted in the merger that forms the remnant is $\sim 7\%$ of the total mass of the merger product, i.e.\ a BNS merger event is clearly capable of producing such a burst.  If the emission is beamed, $E_\mathrm{FRB}$ will be lower by the beaming factor, while if the FRB was detected far from the CHIME beam centre, $E_\mathrm{FRB}$ will be much higher.

\begin{figure}
    \centering
    \includegraphics[width=\textwidth]{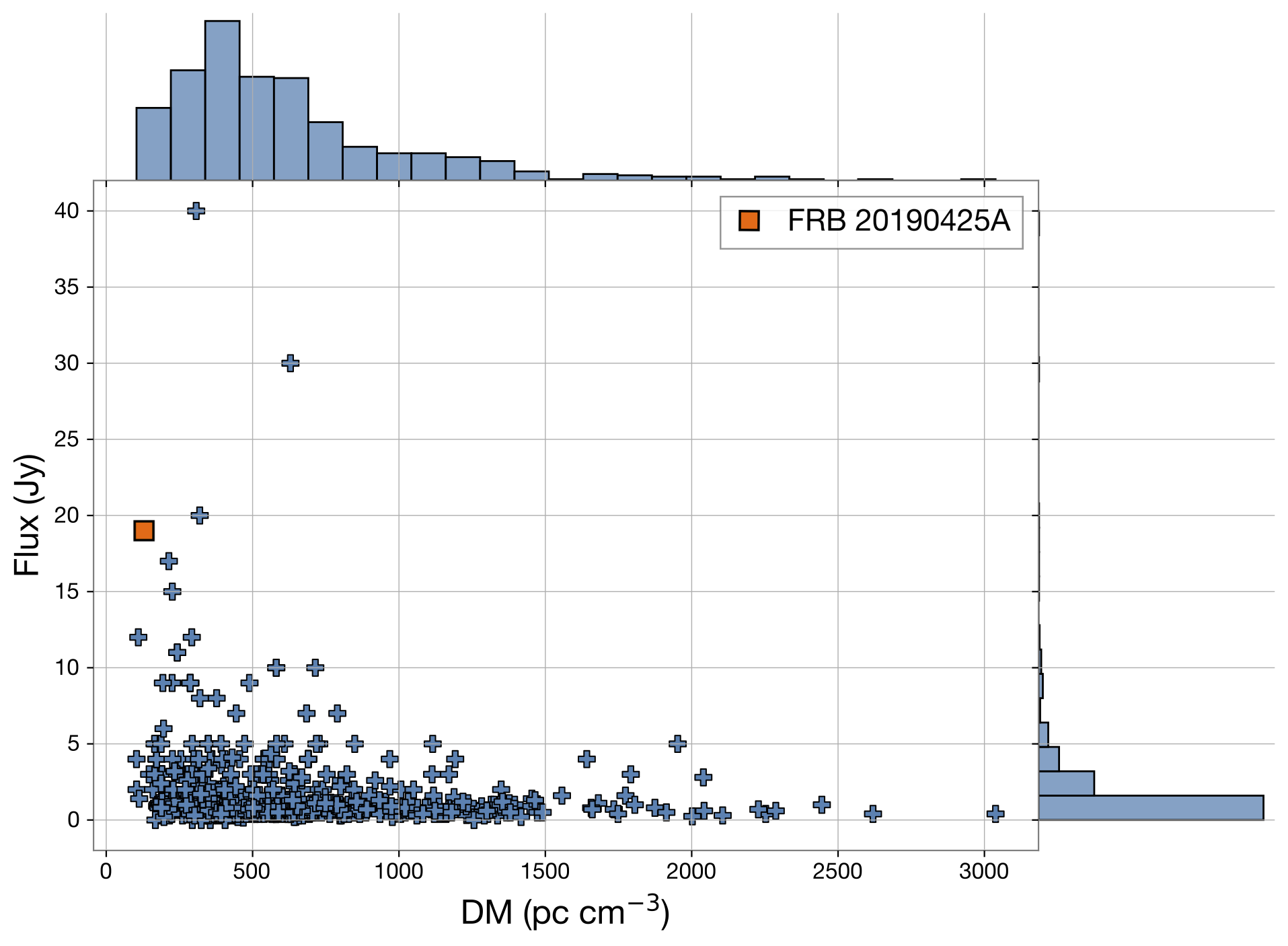}
    \caption{{\bf Flux-DM distribution of CHIME FRBs.} The majority of CHIME FRBs have flux densities below $5\rm \ Jy$. FRB 20190425A (orange square) resides in the low end of the DM spectrum, but has a somewhat exceptional flux density. Note that these flux densities are lower limits, as CHIME flux measurements are derived under the assumption that each burst is detected in the center of the primary beam.}
    \label{fig:fluxdm}
\end{figure}

Here, we also wish to point out the curious brightness of FRB 20190425A. Figure \ref{fig:fluxdm} highlights the flux density distribution of CHIME Catalog 1 FRBs. FRB 20190425A (orange square) exhibits a curiously high flux density for its DM, with only four more notable outliers.

\subsection{GRB association}

Automated electromagnetic follow-up of the real-time detection of GW190425 (S190425z\cite{GCN190425z}) resulted in the reports of a detection of a marginally significant excess of gamma-rays\citemethods{INTEGRALGCN3} by the \textit{INTEGRAL} telescope's SPI-ACS system\citemethods{Savchenko2012}. No significant signal was found by other subsystems on-board \textit{INTEGRAL}, including IBIS/PICsIT, ruling out a localization for the source within the IBIS field of view (FoV). Subsequent analysis, correcting for the local background variance at the time of the event resulted in the updated report of an event detected by SPI-ACS\citemethods{INTEGRALGCN1} with a fluence $F = 2.9\cdot 10^{-10}$ -- $2 \cdot 10^{-9}\,\mathrm{erg\,cm^{2}}$ in the $75-2000\,\mathrm{keV}$ range, six seconds after the GW event, assuming a duration of $1\,\mathrm{s}$. The false alarm probability\cite{INTEGRALGCN2} is less than $3\sigma$ however it cannot be ruled out that the event is physical. We find that the localization provided does not exclude the location of FRB20190425A nor a significant portion of the localization uncertainty for GW190425, however it is noted\cite{INTEGRALGCN2} that the south-west arc of the localization contour is slightly disfavored due to an absence of signal in the IBIS/Veto shield instruments on-board \textit{INTEGRAL}. 
A subsequent independent analysis of \textit{INTEGRAL}/SPI-ACS data by \cite{pozanenko2019} reports the same sGRB event with a fluence of $8\cdot 10^{-8} - 2.4\cdot10^{-6}\,\mathrm{erg\,cm^{-2}}$ with a significance of $5.5 \sigma$, wherein the authors utilize a different calibration model and find an sGRB of significantly longer duration than in the initial prompt analysis\citemethods{INTEGRALGCN1}\cite{ INTEGRALGCN2}\citemethods{INTEGRALGCN3}.

We note that this work is complementary to the on-going international LIGO--Virgo--KAGRA Collaboration
effort to detect sub-threshold prompt GWs associated with CHIME FRBs \citemethods{LVK_CHIME_2022}.

\subsection{Further Discussion}

Capturing the FRB counterparts of a GW source is challenging  given the fact that the CHIME primary beamwidth is
$1.3^\circ$--$2.5^\circ$ in RA, with 200 $\deg^2$ in FoV \cite{chime2021}, covering only around 0.5\% of the sky. What is the chance that an FRB emitted from the post-merger remnant of GW190425 would occur at a time in which it could be detected by CHIME?. The chance of an event emitted at a random time, i.e.\ a random RA, at a declination of $\delta=30^{\circ}$ occurring in this beamwidth is thus ($1.3^\circ$--$2.5^\circ$)/($360^{\circ} \cos \delta )=$0.4\%--0.8\%.
On the other hand, the chance of coincident detection could be improved by several scenarios. 
A bright FRB could be detected far from beam centre, as was the case with a radio flare from SGR 1935+2154, which was observed by CHIME $22^\circ$ from the meridian \cite{chime2020}. This would increase the chance to capture the FRB to $44^{\circ}/(360^{\circ} \cos \delta)$=14\% ($44^\circ$ reflects $22^\circ$ both sides of the beam). Furthermore, there is a slight preference for GWs to be detected over a large sky area around the zenith (and the nadir) of the LIGO detectors. In our case the GW event was detected significantly only by the LIGO-Livingston detector whose zenith aligns well with that of CHIME due to their geographical proximity. Constraining the time interval to 2.5\,hr, while assuming the FRB arises from the same location on the sky, results in a chance detection probability of ($1.3^\circ$--$2.5^\circ$)/($37.5^\circ \cos \delta)=$4\%--8\% for in-beam events, and $22^{\circ}/( 37.5^{\circ} \cos \delta)$=68\% for sidelobe events (only one side of the beam counts, since only detections earlier than 2.5\,hr are considered).

Our favoured interpretation of FRB 20190425A is a blitzar event. Another possibility is that the post-merger magnetar generates multiple repeating bursts and FRB 20190425A is one of them. If this is the case, the likelihood of detecting an FRB associated with the GW source would also have increased. However, we disfavour this possibility for the following reasons: 1. Observationally, FRB 20190425A is bright and carries the key observational signatures of non-repeating FRBs\citemethods{pleunis2021}, e.g. single peak, short duration, and broad spectrum. These are very different from the bursts detected from repeating FRB sources, which typically have long widths, narrow spectra, and very often multiple peaks. 2. Even within the non-repeater population, this burst is brighter than others and is an outlier of the flux-DM relation of most CHIME FRBs (Figure \ref{fig:fluxdm}). This suggests that FRB 20190425A may have a distinct physical origin from other FRBs. 3. Theoretically, interpreting FRB20190425A within the blitzar framework already requires a large $M_{\rm TOV}$ that is marginally consistent with other constraints on the NS EoS. The repeater scenario requires that the post-merger product survive even longer than 2.5 hours, which would further escalate the EoS tension with known results.

The estimated merger rate of BNS (100--1700\,Gpc$^{-3}$\,yr$^{-1}$ \citemethods{LVK_merger_rates_2021}) is far lower than the estimated FRB rate $\sim 10^5$\,Gpc$^{-3}$\,yr$^{-1}$  \cite{ravi2019,luo2020}\citemethods{james2022L}. Thus blitzar FRBs initiated by BNS mergers can account for at most 5\% 
of the FRBs exhibiting broadband, single-peaked morphology observed by CHIME, which themselves account for 30\% of the population. If the association between GW190425 and FRB 20190425A is indeed astrophysical and can be attributed to the blitzar model, it implies the existence of ``subpopulations amongst subpopulations'' of FRBs.

\bigskip
\bigskip
\bigskip
\bibliographystylemethods{naturemag}
\bibliographymethods{methods}

\begin{addendum}
\item 

We acknowledge the custodians of the land this research was conducted on, the Whadjuk (Perth region) Noongar people and pay our respects to elders past, present and emerging. This research has made use of data, software and/or web tools obtained from the Gravitational Wave Open Science Center (\url{https://www.gw-openscience.org/}), a service of LIGO Laboratory, the LIGO Scientific Collaboration and the Virgo Collaboration. LIGO Laboratory and Advanced LIGO are funded by the United States National Science Foundation (NSF) as well as the Science and Technology Facilities Council (STFC) of the United Kingdom, the Max-Planck-Society (MPS), and the State of Niedersachsen/Germany for support of the construction of Advanced LIGO and construction and operation of the GEO600 detector. 
Virgo is funded, through the European Gravitational Observatory (EGO), by the French Centre National de Recherche Scientifique (CNRS), the Italian Istituto Nazionale di Fisica Nucleare (INFN) and the Dutch Nikhef, with contributions by institutions from Belgium, Germany, Greece, Hungary, Ireland, Japan, Monaco, Poland, Portugal, Spain. This research has made use of the NASA/IPAC Extragalactic Database, which is funded by the National Aeronautics and Space Administration and operated by the California Institute of Technology; NASA's Astrophysics Data System Bibliographic Services;  and the Python libraries \textsc{Matplotlib}\cite{Matplotlib2007}, \textsc{NumPy}\cite{Numpy2011}, \textsc{SciPy}\cite{SciPy2019} and \textsc{Pandas}\cite{pandas,reback2020pandas}.
This research has made use of the DSS-2 based on photographic data obtained using The UK Schmidt Telescope. The UK Schmidt Telescope was operated by the Royal Observatory Edinburgh, with funding from the UK Science and Engineering Research Council, until 1988 June, and thereafter by the Anglo-Australian Observatory. The DSS was produced at the Space Telescope Science Institute under US Government grant NAG W-2166.
AM, FHP and MK 
utilized the OzSTAR national facility at Swinburne University of Technology. The OzSTAR program receives funding in part from the Astronomy National Collaborative Research Infrastructure Strategy (NCRIS) allocation provided by the Australian Government. LW, FHP and MK acknowledge funding support  from  Australian Research  Council  Centre  of  Excellence for  Gravitational  Wave  Discovery (OzGrav)  under grant CE170100004. MK acknowledges the SIRF postgraduate scholarship from the University of Western Australia. CWJ acknowledges support from the Australian Government through the Australian Research Council's Discovery Projects funding scheme (project DP210102103). 
SA and BZ acknowledges a Top Tier Doctoral Graduate Research Assistantship (TTDGRA) at University of Nevada, Las Vegas. 

We acknowledge Volodymyr Savchenko and Simon Driver for useful correspondence regarding the sGRB coincidence and galaxy luminosity function respectively, Dr.\ Qi Chu for her knowledge sharing of GW signal extraction, Teresa Slaven-Blair, Tara Murphy, Dougal Dobie, and Hao Qiu for initial discussions relevant to this research, Patrick Sutton for valuable comments regarding the calculation of $P_S$, and Vivek Gupta for information regarding the UTMOST detection of the Crab pulsar.

\item[Author Contributions] 
AM led the GW-FRB coincidence search, GW190425/FRB 20190425A follow-up, chance probability (temporal and spatial) and significance analysis, drafted the initial paper and brought it to completion. LW conceived the original idea for the work, designed the research framework, built the collaboration team, supervised all aspects of the analysis, and contributed to the writing and completion of the paper. CWJ jointly conceived the original idea for the work, contributed to supervision of students in the project, performed the dispersion measure analysis, assisted with significance analysis, and contributed to writing the paper. FHP contributed the host galaxy search, GW parameter estimation, FRB energetics, sGRB context and paper writing and figure showing FRB-host galaxy coincidences. SA and BZ proposed the theoretical interpretation to the data, performed constraints on $M_{\rm TOV}$ for neutron stars and quark stars, and contributed to the writing of the theory part of the paper. MK generated the approximated GW waveform, whitened the public GW strain data and performed matched filtering to construct the SNR time series.

\item[Competing Interests] The authors declare that they have no competing financial interests.

\item[Correspondence] 

 \item[Data Availability Statement]
The CHIME FRB data is publicly available at  \url{https://www.chime-frb.ca/catalog}. The public GW event data is available at \url{https://gracedb.ligo.org/superevents/public/O3/} (general information), \url{https://dcc.ligo.org/LIGO-T1900685/public} (strain data), \url{https://dcc.ligo.org/LIGO-P2000223/public} (GWTC-2 parameter estimation) and \url{https://www.gw-openscience.org/eventapi/html/GWTC-2/GW190425/v2} (GWOSC event portal).
 
 \item[Code Availability Statement]
Processed data is presented in the tables and figures of the paper. Code used for processing data is available upon reasonable requests to the corresponding authors.
 
\end{addendum}

\clearpage

\end{document}